\documentclass{amsart}

\usepackage{setspace}

\usepackage{pifont}

\usepackage[notablist, tablesonly]{endfloat} 

\usepackage{graphicx} 
\usepackage{array} 
\usepackage{bm}
\usepackage{booktabs} 
\usepackage{ctable} 
\usepackage{subfig} 
\usepackage{caption} 
\usepackage{footnote}
\usepackage{threeparttable}
\usepackage{color}
\usepackage{rotating}
\usepackage{lscape}
\usepackage{tensor}
\usepackage[linewidth=1.5pt]{mdframed}
\usepackage{hyperref} 

\usepackage[round]{natbib}
\makeatletter 
\renewcommand\@biblabel[1]{#1.} 
\makeatother %

\graphicspath{{Figures/}} 

\numberwithin{equation}{section}

\DeclareMathOperator{\argmin}{arg\,min}

\DeclareMathOperator{\PP}{\mathbf{P}}

\DeclareMathOperator{\urho}{\underline{\rho}}
\DeclareMathOperator{\uomega}{\underline{\omega}}
\DeclareMathOperator{\usigma}{\underline{\sigma}}
\DeclareMathOperator{\ulambda}{\underline{\lambda}}

\begin{document}

\title{Epitope profiling via mixture modeling \\of ranked data}

\author
[C.       Mollica]
{Cristina Mollica}
\address
{Dipartimento di Scienze statistiche\\
 Sapienza Universit\`a di Roma\\
 Piazzale A. Moro 5\\00185~Roma\\Italy}
\email{cristina.mollica@uniroma1.it}
\author
[L.   Tardella]
{Luca Tardella}
\address
{Dipartimento di Scienze statistiche\\
 Sapienza Universit\`a di Roma\\
 Piazzale A. Moro 5\\00185~Roma\\Italy}
\email{luca.tardella@uniroma1.it}
\thanks
{Version \today} 

\begin{abstract}
We propose the use of probability models for ranked data as a useful alternative to a quantitative data analysis to investigate the outcome of bioassay experiments, when the preliminary  choice of an  appropriate normalization method for the raw numerical responses is difficult or subject to criticism. We review standard distance-based and multistage ranking models and in this last context we propose an original generalization of the \textit{Plackett-Luce model} to account for the order of the ranking elicitation process. The usefulness of the novel model is illustrated 
with its maximum likelihood estimation for a real data set.
Specifically, we address the heterogeneous nature of experimental units via model-based clustering and detail the necessary steps for a successful likelihood maximization through a hybrid version of the Expectation-Maximization algorithm. 
The performance of the mixture model using the new distribution as mixture components is compared with those relative to alternative mixture models for random rankings. A discussion on the interpretation of the identified clusters and a comparison with more standard quantitative approaches are finally provided.
\end{abstract}


\maketitle


\makeatletter \@setaddresses \makeatother \renewcommand{\addresses}{}

\section{Introduction}
\label{s:intro}
Ranked data arise in several contexts, especially when objective and precise measurements of the phenomena of interest can be 
impossible or deemed unreliable and the observer gathers ordinal information in terms of 
orderings,
preferences, judgments, relative or absolute ranking among competitors. 
Research fields where the analysis of ranked data are frequently required 
are the social and behavioral sciences, where studies often ask a sample of $N$ people to rank a finite set of $K$ items according to certain criteria, typically their personal preferences or attitudes. In marketing, political surveys or psychological experiments, items to rank can be consumer goods, political candidates or goals, words or topics considered to be more or less associated to a reference one according to the individual perception. 
Another typical context is sport, where teams, horses or cars compete, 
and the final outcome is a ranking among competitors. 
A detailed and well-structured reference monograph concerning ranking data analysis and modeling is~\citep{Marden}.  

Statistical analysis of observed rankings is less usual in experimental research, 
where the availability of (sometimes sophisticated) measuring devices allows to express phenomena of interest in 
terms of precise quantitative information. 
In this work we verified the usefulness of probability models for ranked data in an experimental study, where quantitative outcomes are indeed available but, for reasons due to numerical instability of the measurements and to the absence of  universally accepted ways of rescaling the original data, one could instead investigate the ranking evidence. For this purpose
we used a real data set from the Large Fragment Phage Display (LFPD) bioassay experiment described in~\citep{Gabrielli:al}. Researchers set up a new promising technology in order to get further insights into the understanding of molecular recognition of the immune system via epitope mapping of the HER2 oncoprotein. They employed a sample of patients and recorded for each subject the binding level, expressed on quantitative scale, between antibodies and specific partially overlapping fragments of the HER2 oncoprotein. The sample was actually composed of three different groups according to the known breast cancer status. A preliminary exploratory analysis of the LFPD data showed differential outcome profiles for the three cancer-specific groups~\citep{Gabrielli:al}. Hence, we assumed the sample as drawn from a heterogeneous, multimodal population and opted to describe it through a mixture modeling approach for the individual ranked binding sequences. 
Two well-studied probability distributions for ranked data, the distance-based and the Plackett-Luce model, and a new extension of the latter have been employed as mixture components and the resulting performances have been compared. Maximum likelihood estimates (MLE) have been obtained with the implementation of the Expectation-Maximization (EM) algorithm or with hybrid versions thereof.

This article is organized as follows: in Section~\ref{s:sect2} we define the notation and review the distance-based and the Plackett-Luce model. The presentation and motivation of our extended Plackett-Luce model follow in Section~\ref{ss:proposal} and the MLE for a finite mixture is discussed in Section~\ref{s:inference}. The application to the LFPD data and the comparison of the novel model with alternative ranking probability distributions are detailed in Section~\ref{s:statana}, with an interpretation of the inferential findings. The article ends with conclusions and proposals for future developments in Section~\ref{concl}.  

\section{Statistical models for ranked data}
\label{s:sect2}

\subsection{Notation and basic definitions}
\label{s:nbd}
Before reviewing some of the approaches for the probabilistic modeling 
of ranked data,
it is convenient to fix some notation. 
Formally, a \textit{full} (or \textit{complete}) \textit{ranking} is a bijective mapping of a finite set $I=\{i_1,\dotsc,i_K\}$ of labeled \textit{items} into a set of \textit{ranks} $R=\{1,\dotsc,K\}$, that is 
$$\pi: I \to R.$$ 
With some abuse of notation, each item label will be identified with its subscript:
instead of writing a ranking as $\pi=(\pi(i_1),\dotsc,\pi(i_K))$, we will simplify it as
$\pi=(\pi(1),\dotsc,\pi(K))$.
In this way positions refer to items and entries give the corresponding assigned ranks, which means that $\pi(i)$ must be read as the rank attributed to the $i$-th item.
The underlying convention is that if $\pi(i)<\pi(i')$, then item $i$ is ranked higher than  item $i'$, and hence preferred to it.

In the literature, one distinguishes 
a \textit{full} from a \textit{partial} (or \textit{incomplete}) \textit{ranking}, in which the rank assignment process is not completely carried out. This happens, for instance, 
when a judge expresses only her first $t$ preferences out of $K$ items 
($t<K$), producing the so-called top-$t$ partial ranking. 
In the present context, the above restrictive definition for ranking is adopted, so that ties are not allowed
due to injectivity 
and partial rankings are not contemplated because of surjectivity of the mapping $\pi$.

The inverse $\pi^{-1}=(\pi^{-1}(1),\dotsc,\pi^{-1}(K))$ of a ranking $\pi$ is called \textit{ordering}. Positions of the components of 
$\pi^{-1}$
refer to ranks and elements correspond to the items.
Hence, $\pi^{-1}(j)$ is the item ranked in the $j$-th position. In order to avoid confusion with
$\pi$, we will henceforth make explicit use of the inverse function notation to denote the corresponding ordering $\pi^{-1}:R\to I$.

We denote with $\mathcal{S}_K$ the set of all $K!$ possible permutations. This special finite subset of $\mathbb{R}^K$ is endowed with a composition operation such that two elements $\pi$ and $\sigma$ in $\mathcal{S}_K$ may yield either a permutation of $R$ or of $I$. In particular, $\pi\sigma^{-1}=\pi\circ\sigma^{-1}=(\pi(\sigma^{-1}(1)),\dotsc,\pi(\sigma^{-1}(K)))$ indicates ranks under $\pi$ of the items ranked $1,\dotsc,K$ by $\sigma$, whereas  $\sigma^{-1}\pi=\sigma^{-1}\circ\pi=(\sigma^{-1}(\pi(1)),\dotsc,\sigma^{-1}(\pi(K)))$ gives items to which $\sigma$ assigns ranks that $\pi$ has attributed to items $1,\dotsc,K$.

When a judge proceeds from the elicitation of her best choice (rank $1$) up to the worst one (rank $K$), we have the so-called \textit{forward ranking process}; the inverse ranking procedure is named \textit{backward ranking process}. This formal definition has been originally introduced in~\citep{Fligner:Verducci-American} but, to our knowledge, the rank assignment scheme has not received an explicit consideration in a model setup in the attempt to improve the description of random ranked data. Obviously, any other order for the rank assignment process is admissible and potentially leads to different models. This aspect has inspired us to expand an existing and well-known parametric ranking model and employ such a new class in the analysis of the LFPD data, in order to 
verify whether and how the reference order can influence the inferential results and the final model-based clustering.

\subsection{Probability models for random rankings}
\label{ss:pmrr}
In this section we give a brief account of rank data modeling. 
For a more systematic review see~\citep{Marden}.
  
The collection 
of all discrete 
distributions for random rankings  
can be identified with the whole $(K!-1)$-dimensional simplex ${\mathcal P}(\mathcal{S}_K)$. This is equivalent to saying that a random ranking and its distribution can be denoted with $\pi \sim P$, where the set $\{P \in {\mathcal P}(\mathcal{S}_K)\}$
%
%
can be regarded as the most general statistical model on rankings 
parameterized by $K!-1$ free parameters, i.e., the probabilities of each ordered sequence. This general form can be considered and named \textit{saturated model} (SM). Within this very general class, a special role is played by the \textit{uniform} or \textit{null model} (UM), represented by the single flat distribution which assigns equal probability to each ranking, and by its opposites, the \textit{degenerate models} (DM), which concentrate all the probability mass on a single ranking. 
Although the SM allows for the maximum degree of flexibility, it becomes intractable and cumbersome to interpret even with a relatively small number $K$ of items, because of the fast-growing dimension of the ranking space. 
These practical limitations have motivated the introduction of simplifying assumptions on the ranking process, in order to deal with subsets of ${\mathcal P}(\mathcal{S}_K)$, and justify the wide assortment of restricted parametric models developed in the rank data theory.

\subsubsection{Distance-based models}
\label{ss:db}

A fundamental class of parametric distributions is the so-called \textit{distance-based model} (DB). Roughly speaking, the DB can be interpreted as the analogue of the normal distribution on the finite discrete space $\mathcal{S}_K$
endowed with the group structure; 
in fact, it is an exponential location-scale model indexed by a discrete parameter $\sigma\in\mathcal{S}_K$, called \textit{modal} or \textit{central ranking}, and a non-negative real \textit{concentration parameter} $\lambda\in\mathbb{R}_+$.
Each distribution in a DB model has the following form
\begin{equation}
\label{e:DB}
P(\pi|\sigma,\lambda)=\frac{1}{Z(\lambda)}e^{-\lambda d(\pi,\sigma)}\qquad\pi\in\mathcal{S}_K,
\end{equation}
where $Z(\lambda)=\sum_{\pi\in\mathcal{S}_K}e^{-\lambda d(\pi,\sigma)}$ is the normalization constant and $d$ is a metric on $\mathcal{S}_K$. The probability mass function in~\eqref{e:DB} attains its maximum at $\pi=\sigma$ and decreases as the distance from $\sigma$ increases.
%
%
Under~\eqref{e:DB} rankings at the same distance from the modal sequence $\sigma$ are equally probable. The central ranking $\sigma$ expresses the so-called \textit{global consensus} in the population,
whereas the concentration/precision parameter $\lambda$ calibrates the effect of $d$ on the probability of the ranking: the higher the value of $\lambda$, the more concentrated the distribution around its mode.
Hence, when $\lambda\to+\infty$, equation~\eqref{e:DB} becomes the DM at $\pi=\sigma$; conversely, when $\lambda=0$ it turns out to be the UM.

Changing the distance measure $d$ in~\eqref{e:DB}, one can 
define
different families of parametric distributions 
for ranked data. 
Formally, a function $d:\mathcal{S}_K\times\mathcal{S}_K\rightarrow\mathbb{R}_+$ 
is a distance between rankings if it 
satisfies the usual three properties of a metric (identity, symmetry and triangle inequality)
%
%
and the additional fourth condition of \textit{right-invariance}
, that is for all $\pi,\xi,\delta\in\mathcal{S}_K$ 
%
\begin{equation}
\label{e:ri}
d(\pi,\xi)=d(\pi\delta^{-1},\xi\delta^{-1}).
\end{equation}
%
Condition~\eqref{e:ri} guarantees the desirable property of invariance of $d$ w.r.t.\ arbitrary relabeling of items. Examples of 
metrics 
for rankings are:
%
\renewcommand{\descriptionlabel}[1]{\textit{#1}}
\begin{description}
\item the \textit{Kendall distance} 
$$d_K(\pi,\xi)={\sum\sum}_{1\leq i<i'\leq K}I_{[(\pi(i)-\pi(i'))(\xi(i)-\xi(i'))<0]},$$
which
counts the number of pairwise disagreements, i.e., the pairs of items with relative discordant order under $\pi$ and $\xi$. It is also equal to the minimum number of adjacent transpositions needed to transform $\pi^{-1}$ into $\xi^{-1}$;
\item the \textit{Spearman distance} $d_{S}(\pi,\xi)=\sum_{i=1}^K[\pi(i)-\xi(i)]^2$
;
\item the \textit{Spearman Footrule} $d_F(\pi,\xi)=\sum_{i=1}^K|\pi(i)-\xi(i)|$;
\item the \textit{Cayley distance} $d_C(\pi,\xi)=K-C(\pi^{-1}\xi)$, where $C(\eta)$ is the number of cycles in $\eta$, corresponding to the minimum number of arbitrary transpositions required to convert $\pi^{-1}$ into $\xi^{-1}$
.
\end{description}
%
%
The reader is referred to~\citep{Critchlow} for a more complete and detailed description of the metrics on rankings.

The computation of $Z(\lambda)$ can be computationally demanding as it requires the summation over all possible rankings. As advised by~\citep{Fligner:Verducci-Royal}, one way to derive a simpler expression for $Z(\lambda)$ is to consider its relation with the moment generating function
of the random variable $D(\pi,\sigma)$ under the UM on $\mathcal{S}_K$.
In the wide variety of distances, only some specific ones lead to a closed form expression for $Z(\lambda)$. Hence, in performing a
statistical  analysis 
of ranked data
one should balance between interpretation purposes, choosing the $d$ which best accommodates the problem at hand, and computational feasibility. For our application to the LFPD data we employed the Kendall distance $d_K$.

\subsubsection{The Plackett-Luce models and related extensions} 
\label{ss:pl}

The \textit{Plackett-Luce model} (PL) is a very popular parametric family for random ranking.
Its name arises from both contributions supplied by~\citep{Luce}, whose monograph provides an in-depth theoretical description of the individual choice behavior with a general axiom, and~\citep{Plackett}, who derived this model in the context of horse races. Its probabilistic expression moves from the decomposition of the ranking process in independent stages, one for each rank that has to be assigned, combined with the underlying assumption of standard forward procedure on the ranking elicitation. In fact, a ranking can be elicited through a series of sequential comparisons in which 
a single item is preferred to all the remaining ones and after being selected is removed from the next comparisons. For this reason, the PL is said to belong to the family of \textit{multistage ranking models}. Specifically, the PL probability distribution is completely specified by the so-called \textit{support parameter} vector $\underline{p}=(p_1,\dotsc,p_K)$, where $p_i>0$ for all $i=1,\dotsc,K$ and $\sum_{i=1}^Kp_i=1$. Note that
in the present formulation the parameters are constrained to add up to one to avoid non-identifiability due to possible multiplication with an arbitrary positive constant. The generic parameter $p_i$ expresses the probability that item $i$ is selected at the first stage of the ranking process and hence preferred among all other items. 
The probability to choose item $i$ at lower preference levels $t>1$ is proportional to its support value $p_i$. Taking into account that the set of available items in the sequence of random selections is reduced 
by one element after each step, the computation of the choice probabilities at each stage for the assignment of the actual rank requires suitable normalization of the support probabilities w.r.t. the set of remaining items at that stage. 
It follows that under the PL the probability of the random ordering $\pi^{-1}$ is
\begin{equation}
\label{e:PL}
\PP(\pi^{-1}|\underline{p})=\prod_{t=1}^{K}\frac{p_{\pi^{-1}(t)}}{\sum_{v=t}^Kp_{\pi^{-1}(v)}}\qquad\pi^{-1}\in\mathcal{S}_K.
\end{equation}
%
%

The vase model metaphor originally introduced by~\citep{Silverberg} is an alternative way to interpret the random stage-wise item selections and a useful representation of the PL to understand its extensions developed in the literature (see~\citep{Marden} for a review). Let us consider a vase containing item-labelled balls such that the vector $\underline{p}$ expresses the starting composition of the vase. The vase differs from an urn simply because the former contains an infinite number of balls in order to allow continuous values of the proportions. At the first stage one draws a ball and ranks the corresponding item first. At the second stage one draws another ball from the vase: if its label is different from $\pi^{-1}(1)$ one assigns rank 2 to the corresponding item, otherwise the ball is put back and one makes drawings until a distinct item is chosen and then ranked second. The multistage experiment ends when there is only one item not yet selected and this is automatically ranked last. The probability of a generic sequence of drawings turns out to be~\eqref{e:PL}. In such a scheme the vase configuration is constant over all stages and interactions among items are not contemplated. A first attempt to generalize this basic scheme consists in retaining the absence of item interactions but letting the vase composition vary among stages, as formalized in~\citep{Silverberg}. In this model setting the support parameters become stage-dependent, that is $p_{ti}$ for $t=1,\dots,(K-1)$ and $i=1,\dots,K$.
Setting the special form $p_{ti}=p_i^{\alpha_t}$ one obtains the \textit{Benter model} (BM) introduced by~\citep{Benter},
%
%
where the parameter vector $\underline{\alpha}=(\alpha_1,\dotsc,\alpha_K)$ with $0\leq\alpha_t\leq1$ for all $t=1,\dotsc,K$ is named \textit{dampening parameter} and accommodates for the possible different degree of accuracy the choice at each selection stage is made with. The PL corresponds to the BM with $\alpha_t=\alpha=1$ for all $t=1,\dotsc,K$
.
Relaxing also the non-interaction hypothesis, meaning that the vase composition at each stage relies on the previous selected items,~\citep{Plackett} defined a hierarchy of further extensions of the PL. They are referred to in~\citep{Marden} as \textit{Lag $L$ models}, where $L=0,\dotsc,K-2$ indicates that the vase at stage $t$ depends on the previous choices only through the last $L$ selected items \{$\pi^{-1}(t-L),\dotsc,\pi^{-1}(t-1)\}$. The Lag 0 model
 coincides with the ordinary PL.
The Lag 1 model is such that at each choice step $t$ the vase depends only on the item $\pi^{-1}(t-1)$.
In general, the total number of parameters in the Lag $L$ model is given by
%
%
$K(K-1)\cdots(K-L)-1$, thus 
the $L=K-2$ model corresponds to the SM.

\subsection{Novel extension of the Plackett-Luce model}
\label{ss:proposal}

In this section we introduce an original proposal to generalize the standard PL. Multistage ranking models previously reviewed implicitly suppose that preferences are expressed with the canonical forward procedure, proceeding with the assignment of the first rank up to the last one. This is just a preliminary assumption and other reference orders can be contemplated but, to our knowledge, this aspect has not been addressed in the literature.  Indeed, even the individual experience in choice problems suggests the plausibility of alternative paths for the ranking elicitation. For example, one can think of situations where the judge has a clearer perception about her most- and least-liked items first but only a vaguer idea relative to middle ranks; alternatively again the ranker can build up her best alternatives following an exclusion process starting with the final position, which would be described by a backward model. Besides the motivation to characterize typical behaviors in real choice/selection problems, we can also aim at obtaining a more flexible tool in order to improve the description of observed phenomena collected in the form of ordered data. All these intuitive and practical instances make the forward hypothesis too restrictive when approaching a flexible inferential analysis of a ranking data set. Hence, we propose to extend the PL in this way: rather than fixing \textit{a priori} the stepwise order leading the judge to her final
ranked sequence, we would like to 
represent it with a specific free parameter $\rho\in\mathcal{S}_K$ in the model and let data guide inference about the reference order followed in the rank assignment scheme. Such an approach would also alleviate the asymmetry toward ranks assigned at the extreme (the first and the last) stages of the ranking procedure, which by nature affects the PL with hypothesized known reference order. It turns out that the reference order $\rho=(\rho(1),\dotsc,\rho(K))$ is the result of a bijection between the stage set $S$ and the rank set $R$, i.e.,
\begin{equation*}
\rho:S\to R,
\end{equation*}
where the entry $\rho(t)$ indicates the rank attributed at the $t$-th stage of the ranking process. Then, $\rho$ identifies a discrete parameter taking values in $\mathcal{S}_K$. The composition of an ordering $\pi^{-1}$ with a reference order $\rho$ yields
\begin{equation*}
\label{e:eta}
\eta^{-1}=\pi^{-1}\rho,
\end{equation*}
the sequence listing the items selected at each stage. This means that $\eta^{-1}(t)=\pi^{-1}(\rho(t))$ is the item chosen at step $t$ and receiving rank $\rho(t)$. The probability of a random ordering under the \textit{extended Plackett-Luce model} can be written as
\begin{equation}
\label{e:EPL}
\PP_{EPL}(\pi^{-1}|\rho,\underline{p})=\PP_{PL}(\pi^{-1}\rho|\underline{p})
=\prod_{t=1}^K\frac{p_{\pi^{-1}(\rho(t))}}{\sum_{v=t}^Kp_{\pi^{-1}(\rho(v))}}\qquad\pi^{-1}\in\mathcal{S}_K,
\end{equation}
where the additional discrete parameter $\rho$ acts directly on the right of the generated outcome of a standard PL.
Hereafter we will shortly refer to~\eqref{e:EPL} as EPL($\rho$,$\underline{p}$). The vector $\underline{p}$ continues to denote the support parameters with the probabilities for each item to be selected at the first stage and receiving rank given by the first entry in $\rho$. Obviously, the standard PL is a special case of the EPL, obtained setting $\rho$ equal to the identity permutation $e=(1,2,\dots,K)$. Similarly, when $\rho=(K+1)-e$ one has the backward PL.

From a theoretical point of view, \eqref{e:EPL} is a proper generalization of the \eqref{e:PL} if and only if such a new class covers a wider portion of the SM, i.e., if the novel EPL allows to describe additional probability functions that can not be derived with any parameter specification from the PL. In other words, one should give a formal proof concerning the existence of a ranking distribution, generated by the new EPL, which does not belong to the standard PL family. Such a proof is given in the Appendix. In section~\ref{ss:EPLest} we describe in detail the MLE of such a new model.

\subsection{Finite mixture modeling for ranked data}

One of the 
nice formal 
properties satisfied by the DB~\eqref{e:DB} is strong unimodality, meaning that the probability decreases as the distance from the modal ranking increases,
see~\citep{Marden}. 
On the other hand
strong unimodality
is expected to be violated in
real data, especially when the sample composition is heterogenous 
w.r.t. 
factors related to the ranking elicitation.
A well-established statistical tool to address inference in the  presence of unobserved heterogeneity is given by the finite mixture approach. A \textit{finite mixture model} assumes that the population consists of a finite number $G$ of subpopulations. In this setting the probability of observing the ranking $\pi_s$ for the $s$-th unit is
\begin{equation*}
f(\pi_s)=\sum_{g=1}^G\omega_gf_g(\pi_s)\qquad\pi_s\in\mathcal{S}_K,
\end{equation*}
where $f_g(\cdot)$ denotes the $g$-th \textit{component} of the mixture, i.e., the statistical distribution of data within the $g$-th group and $\omega_g$ is the probability for the $s$-th observation to belong to the $g$-th group.
The membership probabilities $\underline\omega=(\omega_1,\cdots,\omega_G)$ are usually termed \textit{weights} of the mixture components.  
Mixture components are often modeled with members of the same parametric family, that is, $f_g(\cdot)=f(\cdot|\theta_g)\in\{f(\cdot|\theta): \theta\in\Theta\}$ for all $g=1,\dotsc,G$ and thus they are identified by the group-specific parameter $\theta_g$. For a more  extensive introduction to finite mixture models the reader can refer to~\citep{McLach:Peel}.
In the ranking literature one can find several recent mixture model
applications to 
make the ranked data modeling more flexible and 
account for unobserved heterogeneity. 
For example,~\citep{Murphy:Martin} analyzed the popular 1980 APA (American Psychological Association) presidential election data set, in particular the sub-data set of complete rankings, with a mixture of distance-based models. 
They aimed at inquiring 
voters' orientation towards candidates within the electorate, assessing the possible adequacy to incorporate a noise component (UM) in the mixture.
Such a component, in fact, could collect outliers and/or observations characterized by untypical preference profiles with a possible final improvement of model fitting. 
A similar method was adopted in other preference studies.~\citep{Gormley:Murphy-Royal} fitted a mixture of PL to the 2000 CAO (Central Applicant Office) data set to investigate motivations driving Irish college applicants in the degree course choice. ~\citep{Gormley:Murphy-American} applied a mixtures of both PL and BM to infer the structure of the Irish political electorate and characterize voting blocks. In subsequent works the same authors attempted to extend such an approach in different directions (for further details see~\citep{Gormley:Murphy-AnnalsApplied} and~\citep{Gormley:Murphy-BayesianAnalysis}).

In section~\ref{LFPDana} we present our application of mixture models 
for ranking to data 
originated 
from the LFPD bioassay experiment. 
For the analysis of the LFPD data set we
implemented different mixture models, 
adopting as mixture components elements from the following parametric families:
\begin{itemize}
\item DB with $d=d_K$;
\item PL with known forward and backward reference order;
\item our novel EPL.
\end{itemize}
DB and PL represent two of the most frequently used distributions for inferring ranking data and both parameterizations allow a clear interpretation: in the former, the central ranking summarizes the 
profile
of the population in assessing the orderings of the items and the concentration parameter expresses how representative the modal ranking is; in the latter, the higher the item support parameter value, the greater the probability for that item to be preferred at each selection level.
For the argument on the choice of the EPL in the analysis of the LFPD data, the reader is referred to section~\ref{LFPDana}.

\section{Inferring ranking models}
\label{s:inference}

\subsection{MLE of the mixture of distance-based models}

We briefly summarize here the fundamental steps to derive the MLE for a mixture of DB with $d=d_K$ when a sample $\underline\pi=(\pi_1,\dotsc,\pi_s,\dotsc,\pi_N)$ is available. We basically reproduce the algorithm described in~\citep{Murphy:Martin}.
Let $\underline{z}_s=(z_{s1},\dotsc,z_{sG})$ be the latent variable indicating the individual component membership such that
\begin{equation*}
z_{sg}=
\begin{cases}
1\qquad\text{if the $s$-th unit belongs to the $g$-th group }, \\
0\qquad\text{otherwise},
\end{cases}
\end{equation*}
for $s=1,\dotsc,N$.
From~\eqref{e:DB} it follows that 
the complete log-likelihood can be written as
\begin{equation*}
\label{e:mDB}
l_C(\usigma,\ulambda,\uomega,\underline{z}) 
=\sum_{s=1}^N\sum_{g=1}^Gz_{sg}[\log\omega_g-\lambda_gd_K(\pi_s,\sigma_g)-\log Z(\lambda_g)],
\end{equation*}
where $\uomega$ and $\ulambda$ are vectors representing, respectively, the prior group membership probabilities and the component-specific concentration parameters, whereas $\usigma$ is a $G\times K$ matrix, whose rows indicate the central rankings of the mixture components. To derive parameter estimates the EM algorithm can be implemented; it represents the major scheme to address the inferential analysis in the presence of missing data \citep{Demp:Lai:Rub}. 
For the present model the EM algorithm consists of the following steps:
\begin{itemize}
\item[Initialization:]
set initial values $\usigma^{(0)},\ulambda^{(0)},\uomega^{(0)}$ for the parameters to be estimated (we used random starting values). 
\item[E-step:\phantom{mmmi}] 
at iteration $l+1$ 
compute
\begin{equation*}
\label{e:EstepDB}
\hat z_{sg}^{(l+1)}=\frac{\omega_g^{(l)}\PP_{DB}(\pi_s|\sigma_g^{(l)},\lambda_g^{(l)})}{\sum_{g'=1}^G\omega_{g'}^{(l)}\PP_{DB}(\pi_s|\sigma_{g'}^{(l)},\lambda_{g'}^{(l)})},
\end{equation*}
for $s=1,\dotsc,N$ and $g=1,\dotsc,G$,
which is the current estimate of the posterior probability that subject $s$ belongs to the $g$-th component;
\item[M-step:\phantom{mmm}]
at iteration $l+1$ 
compute
\begin{align*}
\omega_g^{(l+1)}&=\sum_{s=1}^N\frac{\hat z_{sg}^{(l+1)}}{N},\\
\sigma_g^{(l+1)}&=\underset{\sigma}{\argmin}\sum_{s=1}^N\hat z_{sg}^{(l+1)}d(\pi_s,\sigma),
\end{align*}
and determine $\lambda_g^{(l+1)}$ as the solution of
\begin{equation*}
\frac{Ke^{-\lambda}}{1-e^{-\lambda}}-\sum_{j=1}^K\frac{je^{-j\lambda}}{1-e^{-j\lambda}}=\frac{\sum_{s=1}^N\hat z_{sg}^{(l+1)}d(\pi_s,\sigma_g^{(l+1)})}{\sum_{s=1}^N\hat z_{sg}^{(l+1)}},
\end{equation*}
\end{itemize}  
for $g=1,\dotsc,G$. 
We run the algorithm 
with a suitably large number of different starting values to 
address the issue of 
local maxima.

\subsection{MLE of the mixture of Extended Plackett-Luce models}
\label{ss:EPLest}

As mentioned before, the conventional forward PL is a reduction of the wider family of EPL distributions obtained setting the reference order parameter $\rho$ equal to the identity permutation $e$. It follows that the estimation procedure for the mixture of PL can be easily deduced from the one derived for the mixture of EPL with all known reference orders $\rho_g=e$. However, explicit estimation formulas for this special case can be found in~\citep{Gormley:Murphy-Royal}. In this section we restrict ourselves to give inferential details only for the 
extended 
model, starting with the simpler case of homogenous population (G=1).

Postulating the EPL($\rho$,$\underline{p}$) as the underlying mechanism generating the observed orderings $\underline{\pi}^{-1}=(\pi_1^{-1},\dotsc,\pi_N^{-1})$,
the log-likelihood has the following expression
\begin{equation}
\label{e:Log.lik.EPL}
\begin{split}
l(\rho,\underline{p}) 
&=\sum_{s=1}^N\sum_{t=1}^K\biggl[\log\frac{p_{\pi_s^{-1}(\rho(t))}}{\sum_{v=t}^Kp_{\pi_s^{-1}(\rho(v))}}\biggr]\\
&=\sum_{s=1}^N\sum_{t=1}^K\biggl[\log p_{\pi_s^{-1}(\rho(t))}-\log\biggl(\sum_{v=t}^Kp_{\pi_s^{-1}(\rho(v))}\biggr)\biggr]\\
&=N\sum_{i=1}^K\log p_i-\sum_{s=1}^N\sum_{t=1}^K\log\biggl(\sum_{v=t}^Kp_{\pi_s^{-1}(\rho(v))}\biggr).
\end{split}
\end{equation}
Note that in order to find MLE solutions, the direct maximization of the log-likelihood w.r.t. the $p$'s is made arduous by the presence of the annoying term $\log\biggl(\sum_{v=t}^Kp_{\pi_s^{-1}(\rho(v))}\biggr)$. So, we derived the estimation formula for the support parameters borrowing the approach detailed in~\citep{Hunter} and based on the Minorization/Maximization (MM) algorithm. Such an iterative optimization method is reviewed in general in~\citep{Lange:Hunter:Yang} and~\citep{Hunter:Lange}, whereas~\citep{Hunter} discusses the specific application of the MM algorithm for the estimation of the PL. The basic idea consists of performing the optimization step for the $p$'s on a surrogate objective function rather than on~\eqref{e:Log.lik.EPL}. The surrogate is obtained by exploiting the strict convexity of $-\log\biggl(\sum_{v=t}^Kp_{\pi_s^{-1}(\rho(v))}\biggr)$ and in particular the supporting hyperplane property for convex functions. From Taylor's linear expansion, in fact, one has
\begin{equation*}
\label{}
-\log\left(\sum_{v=t}^Kp_{\pi_s^{-1}(\rho(v))}\right)\geq1-\log\left(\sum_{v=t}^Kp_{\pi_s^{-1}(\rho(v))}^{(l)}\right)-\sum_{v=t}^Kp_{\pi_s^{-1}(\rho(v))}/\sum_{v=t}^Kp_{\pi_s^{-1}(\rho(v))}^{(l)},
\end{equation*}
and disregarding terms not depending on $\underline{p}$ the minorizing auxiliary objective function can be written as
\begin{equation}
\label{e:q}
q=N\sum_{i=1}^K\log p_i-\sum_{s=1}^N\sum_{t=1}^K
\frac{\sum_{v=t}^Kp_{\pi_s^{-1}(\rho(v))}}{\sum_{v=t}^Kp_{\pi_s^{-1}(\rho(v))}^{(l)}}.
\end{equation}
As emphasized by~\citep{Hunter}, the convenience of optimizing the more tractable~\eqref{e:q} in place of~\eqref{e:Log.lik.EPL} is the possibility to  estimate each support parameter $p_i$ separately. 
Furthermore, the iterative maximization of $q$ returns a sequence $\underline{p}^{(1)},\underline{p}^{(2)},\dotsc$ that is guaranteed to converge at least to a local maximum of the original objective function. 
Thus, 
we can differentiate w.r.t. each $p_i$ and get 
\begin{equation}
\label{e:diff}
\frac{\partial q}{\partial p_i}=\frac{N}{p_i}-\sum_{s=1}^N\sum_{t=1}^K
\frac{\delta_{sti}^{(l)}}{\sum_{v=t}^Kp_{\pi_s^{-1}(\rho(v))}^{(l)}},
\end{equation}
where
\begin{equation*}
\delta_{sti}^{(l)}=
\begin{cases}
      1\qquad\text{ if }i\in\{\pi_s^{-1}(\rho^{(l)}(t)),\cdots,\pi_s^{-1}(\rho^{(l)}(K))\},\\
      0\qquad\text{ otherwise},
\end{cases}
\end{equation*}
corresponds to the binary indicator for the event that item $i$ is still available at stage $t$ 
or, equivalently, that is not selected by unit $s$ before stage $t$. Notice that the binary array has a superscript because of the dependence on the $\rho=\rho^{(l)}$ available at the current iteration. Equating~\eqref{e:diff} to zero, the updating rule at the current iteration for $p_i$ is
\begin{equation*}
\label{e:suppest1}
p^{(l+1)}_i=\frac{N}{\sum_{s=1}^N\sum_{t=1}^K
\frac{\delta_{sti}^{(l)}}{\sum_{v=t}^Kp_{\pi_s^{-1}(\rho^{(l)}(v))}^{(l)}}}\qquad i=1,\cdots,K.
\end{equation*}
The update of the reference order parameter is obtained using the original log-likelihood as follows
\begin{equation}
\label{e:refest}
\rho^{(l+1)}=\underset{\rho}{\argmin}\sum_{s=1}^N\sum_{t=1}^K\log\biggl(\sum_{v=t}^Kp_{\pi_s^{-1}(\rho(v))}^{(l+1)}\biggr).
\end{equation}
Solving~\eqref{e:refest} with a global search in $\mathcal{S}_K$ is prohibitive when $K$ is large, as in our application to the LFPD data. 
So, we implemented 
a local search similarly to 
the method suggested by~\citep{Busse:al} and constrained the optimization within a fixed Kendall distance from the current estimate for the reference order $\rho^{(l)}$.

Now we relax the hypothesis of homogeneous population and consider a more flexible mixture model with EPL components, discussing the related inferential issues. If we assume our random sample $\underline{\pi}^{-1}$ drawn from a mixture of EPL, the probability of the generic ordering is written as the average of its probability under each sub-population weighted with the corresponding mixture component weight, i.e.,
\begin{equation*}
\begin{split}
\label{e:mixEPL}
\PP(\pi_s^{-1}|\urho,\underline{p},\uomega)&=\sum_{g=1}^G\omega_g\PP_{EPL}(\pi_s^{-1}|\rho_g,\underline{p}_g)\\
&=\sum_{g=1}^G\omega_g\prod_{t=1}^K\frac{p_{g\pi_s^{-1}(\rho_g(t))}}{\sum_{v=t}^Kp_{g\pi_s^{-1}(\rho_g(v))}}.
\end{split}
\end{equation*}
Hence, the observed log-likelihood turns out to be 
\begin{equation*}
\label{e:Log.lik.mixEPL}
l(\urho,\underline{p},\uomega)=\sum_{s=1}^N\log\biggl[\sum_{g=1}^G\omega_g\prod_{t=1}^K\frac{p_{g\pi_s^{-1}(\rho_g(t))}}{\sum_{v=t}^Kp_{g\pi_s^{-1}(\rho_g(v))}}\biggr].
\end{equation*}
Augmenting data with the missing individual group membership indicator $\underline{z}_s=(z_{s1},\cdots,z_{sG})$,
one obtains the following expression for the complete log-likelihood   
\begin{equation*}
\begin{split}
\label{e:CLog.lik.mixEPL}
l_C(\urho,\underline{p},\uomega,\underline{z})&=\log \PP(\underline{\pi}^{-1},\underline{z}|\urho,\underline{p},\uomega)
=\log\prod_{s=1}^N\PP(\pi^{-1}_s|\underline{z}_s,\urho,\underline{p})\PP(\underline{z}_s|\uomega)\\
&=\log\prod_{s=1}^N\prod_{g=1}^G\biggl[\omega_g\prod_{t=1}^K\frac{p_{g\pi_s^{-1}(\rho_g(t))}}{\sum_{v=t}^Kp_{g\pi_s^{-1}(\rho_g(v))}}\biggr]^{z_{sg}}\\
&=\sum_{s=1}^N\sum_{g=1}^Gz_{sg}\biggl[\log\omega_g+\sum_{i=1}^K\log{p_{gi}}-\sum_{t=1}^K\log\left({\sum_{v=t}^Kp_{g\pi_s^{-1}(\rho_g(v))}}\right)\biggr].
\end{split}
\end{equation*}
In the EM algorithm the maximization problem is transferred on the the expectation of the $l_C$ w.r.t. the posterior distribution of the latent variables $\underline{z}$ represented by $\underline{\hat{z}}$, that is
\begin{equation*}
Q=\mathbf{E}[l_C|\underline{\pi}^{-1},\urho,\underline{p},\uomega]=\sum_{s=1}^N\sum_{g=1}^G\hat z_{sg}\biggl[\log\omega_g+\sum_{i=1}^K\log{p_{gi}}-\sum_{t=1}^K\log\left({\sum_{v=t}^Kp_{g\pi_s^{-1}(\rho_g(v))}}\right)\biggr],
\end{equation*}
where
\begin{equation*}
\label{e:EstepPL}
\hat z_{sg}^{(l+1)}=\frac{\omega_g^{(l)}\PP_{EPL}(\pi^{-1}_s|\rho_g^{(l)},\underline{p}_g^{(l)})}{\sum_{g'=1}^G\omega_{g'}^{(l)}\PP_{EPL}(\pi^{-1}_s|\rho_{g'}^{(l)},\underline{p}_{g'}^{(l)})},
\end{equation*}
for $s=1,\dotsc,N$ and $g=1,\dotsc,G$.
Similarly to \citep{Gormley:Murphy-Royal} we combined the EM with the MM algorithm into a hybrid version of the former called EMM algorithm using the following minorizing surrogate function
\begin{equation*}
Q\geq q=\sum_{s=1}^N\sum_{g=1}^G\hat z_{sg}\sum_{i=1}^K\log{p_{gi}}-
\sum_{s=1}^N\sum_{g=1}^G\hat z_{sg}\sum_{t=1}^K
\frac{\sum_{v=t}^Kp_{g\pi_s^{-1}(\rho_g(v))}}{\sum_{v=t}^Kp_{g\pi_s^{-1}(\rho_g(v))}^{(l)}}.
\end{equation*}
Thus, differentiating
\begin{equation}
\label{e:different}
\frac{\partial q}{\partial p_{gi}}=\frac{\sum_{s=1}^N\hat z_{sg}}{p_{gi}}-
\sum_{s=1}^N\hat z_{sg}\sum_{t=1}^K
\frac{\delta_{stig}}{\sum_{v=t}^Kp_{g\pi_s^{-1}(\rho_g(v))}^{(l)}}
\end{equation}
and equating~\eqref{e:different} to zero, the updating rule for $p_{gi}$ at the current iteration is
\begin{equation*}
\label{e:suppest2}
p_{gi}^{(l+1)}=\frac{\sum_{s=1}^N\hat z_{sg}^{(l+1)}}{\sum_{s=1}^N\hat z_{sg}^{(l+1)}\sum_{t=1}^K
\frac{\delta_{stig}^{(l)}}{\sum_{v=t}^Kp_{g\pi_s^{-1}(\rho_g^{(l)}(v))}^{(l)}}}
\end{equation*}
for $g=1,\cdots,G$ and $i=1,\cdots,K$ with
\begin{equation*}
\delta_{stig}^{(l)}=
\begin{cases}
      1\qquad\text{ if }i\in\{\pi_s^{-1}(\rho_g^{(l)}(t)),\cdots,\pi_s^{-1}(\rho_g^{(l)}(K))\},\\
      0\qquad\text{ otherwise},
\end{cases}
\end{equation*}
indicating if, under the 
group-specific reference order
$\rho_g$, the unit $s$ has not extracted the $i$-th item before stage $t$, and hence  if at that step it still belongs to the set of available alternatives or not. The update for the reference orders for each subgroup is  
\begin{equation*}
\label{e:refestmix}
\rho_g^{(l+1)}=\underset{\rho}{\argmin}\sum_{s=1}^N\hat{z}_{sg}^{(l+1)}\sum_{t=1}^K\log\left(\sum_{v=t}^Kp_{g\pi_s^{-1}(\rho(v))}^{(l+1)}\right)\qquad g=1,\cdots,G.
\end{equation*}
As in the case $G=1$ in~\eqref{e:refest}, the above minimization is performed locally. The M-step ends with the update of the mixture weights, computed as the posterior proportions of sample units belonging to each group
\begin{equation*}
\omega_g^{(l+1)}=\frac{\sum_{s=1}^N\hat z_{sg}^{(l+1)}}{N}\qquad g=1,\cdots,G.
\end{equation*}
%


\subsection{Algorithm convergence and model selection}

We conducted MLE inference 
for DB and EPL mixture models 
relying 
on the EM algorithm and on a hybrid version thereof. We developed a suite of functions written in R language~\citep{Rsoft} which are available upon request from the first author. 
In these estimation procedures the log-likelihood is iteratively maximized until convergence is achieved. As suggested by~\citep{McLach:Peel}, 
the Aitken acceleration criterion 
has been employed to assess convergence, 
in place of 
the standard lack of progress criterion based on the absolute/relative increment of the log-likelihood.
For a discussion on the relative merits of the Aitken acceleration criterion and other related proposals, see~\citep{McNich:Murphy}. 

Another crucial issue in a mixture modeling setting is the choice of the number of components. In the statistical literature this problem is addressed with several criteria; we opted for the popular \textit{Bayesian Information Criterion} 
\begin{equation*}
\text{BIC}=-2l(\hat\theta_{ML})+\nu\log N,
\end{equation*}
where $l(\hat\theta_{ML})$ is the maximized log-likelihood and $\nu$ is the number of free parameters. The BIC, introduced by~\citep{Schwarz}, is a measure 
which 
balances 
between two 
conflicting goals
typically 
aimed at when fitting
a statistical model: good fit
and parameter parsimony,
where the latter is 
modulated through
the penalty term. 
In the presence of competing mixture models, the one associated with the lowest value of the BIC is preferred.
In the next section we detail MLE results derived from alternative mixture models fitted to the LFPD data set.

\section{Statistical analysis of LFPD data}
\label{s:statana}

\subsection{The LFPD data set}

\begin{figure}
\begin{center}
\includegraphics[scale=0.55]{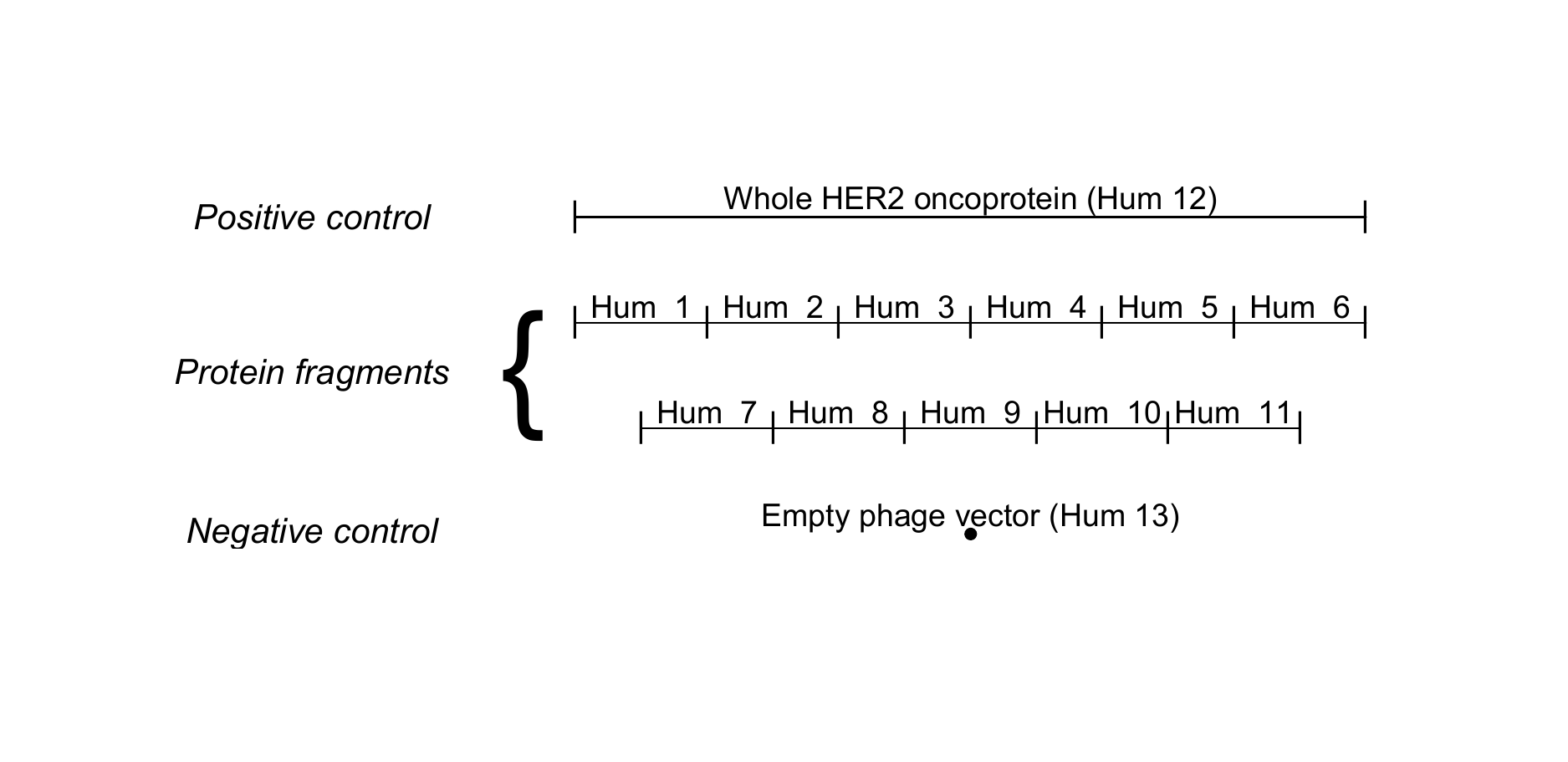}
\caption
{1-D scheme of the HER2 oncoprotein and its segmentation into 11 partially overlapping fragments (Hum) employed in the LFPD bioassay experiment. Hum 12 and Hum 13 indicate respectively the whole HER2 oncoprotein (positive control) and the empty phage vector (negative control).}
\label{HER2}
\end{center}
\end{figure}

Our investigation is motivated by a real
data set coming from a new technology for epitope mapping of the binding between the antibodies present in a biological tissue and a target protein. The biological foundation of the experiment is detailed in~\citep{Gabrielli:al} and consists of repeated binding measurements of human blood exposed to $K=11$ partially overlapping fragments of the HER2 oncoprotein, denoted sequentially by Hum 1,$\cdots$, Hum 11 (see~Figure~\ref{HER2}). Researchers were originally interested in testing the validity of 
their  
innovative biotechnology which consists in a new way of isolating protein fragments without losing the conformational structure of the protein portions. To achieve this goal they employed a phage as a vector for hosting each protein fragment. Then they compared the binding outcome detected on each of the 11 fragments via a standard Enzyme-Linked Immunosorbent Assay (ELISA) with the whole protein (Hum 12) and the empty vector (Hum 13), used respectively as positive and negative controls (see~Figure~\ref{HER2}).
They first checked with monoclonal antibodies that the expected binding at some specific fragment was actually detected. Then they 
gathered 
$N=67$ samples of human blood taken from three different disease groups: i) HD $=$ healthy patients, ii) EBC $=$ patients diagnosed with breast cancer at an early stage, iii) MBC $=$ patients diagnosed with metastatic breast cancer. Binding outcomes from the ELISA experiment have been detected by a laser scanner so that 
the binding intensities have been measured and recorded 
in terms of absorbance levels in nanometers (nm). 
In the next section we motivate our statistical analysis of the LFPD data based on the ordinal information, rather than on the original quantitative scale measurements. 

\subsection{Mixture models for the LFPD data}
\label{LFPDana}

\begin{figure}
\begin{center}
\includegraphics[scale=0.5]{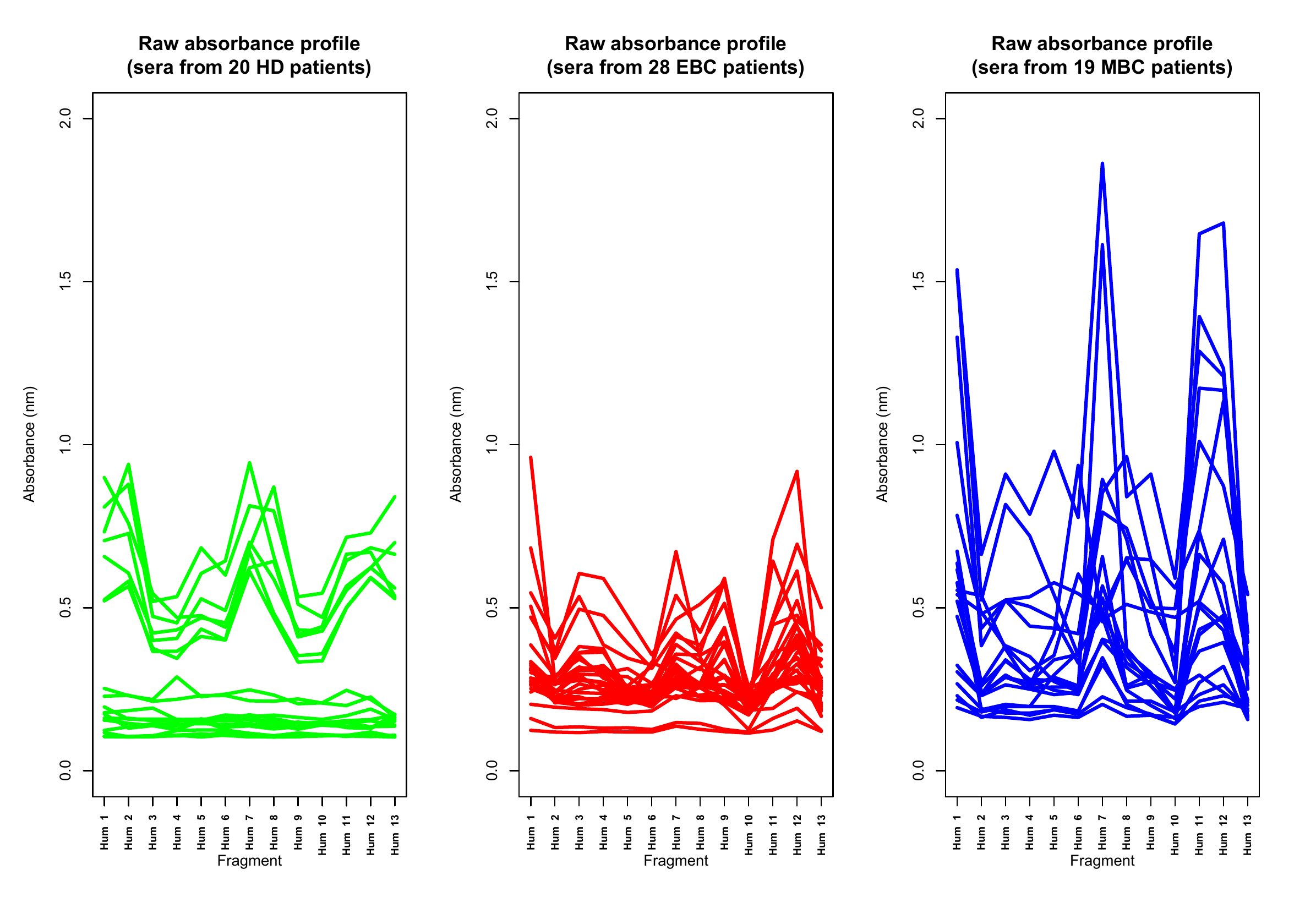}
\caption
{Raw absorbance profiles for the three group of patients in the LFPD study:
HD  $=$ healthy (green),
EBC $=$ diagnosed with breast cancer in an early stage (red),
MBC $=$ diagnosed with metastatic breast cancer (blue).
Each broken line represents the absorbance levels in the HER2 oncoprotein fragments (Hum) of a single experimental unit. Hum 12 and Hum 13 indicate respectively the whole HER2 oncoprotein (positive control) and the empty phage vector (negative control).}
\label{Raw}
\end{center}
\end{figure}

\begin{figure}
\begin{center}
\includegraphics[scale=0.5]{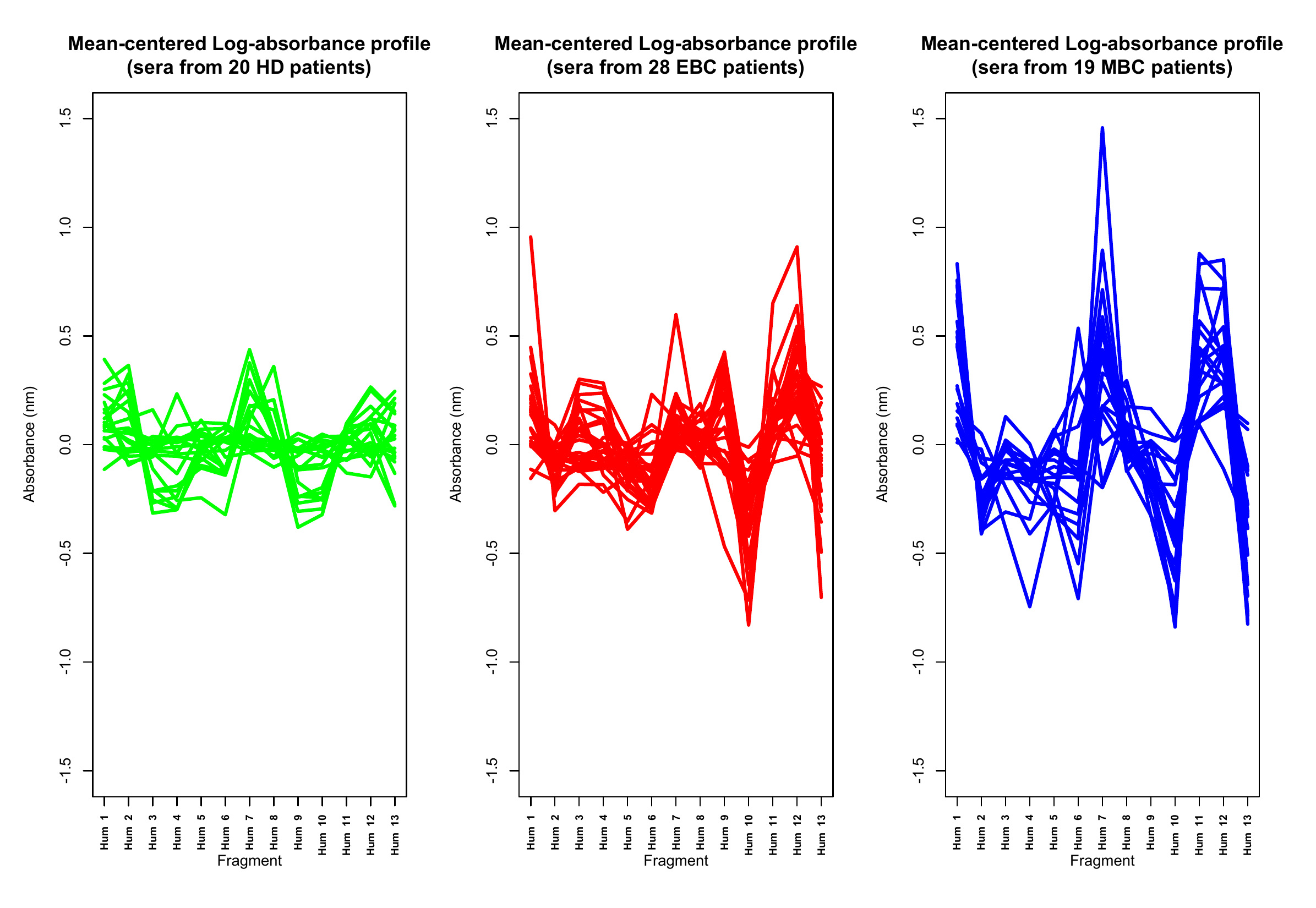}
\caption
{Mean-centered log-absorbance profiles for the three group of patients in the LFPD study:
HD  $=$ healthy (green),
EBC $=$ diagnosed with breast cancer in an early stage (red),
MBC $=$ diagnosed with metastatic breast cancer (blue).
Each broken line represents the mean-centered log-absorbance levels in the HER2 oncoprotein fragments (Hum) of a single experimental unit. Hum 12 and Hum 13 indicate respectively the whole HER2 oncoprotein (positive control) and the empty phage vector (negative control).}
\label{LogCen}
\end{center}
\end{figure}

The original raw absorbance data derived from the LFPD experiment were somehow wildly fluctuating 
and looked indeed very heterogeneous as apparent in~Figure~\ref{Raw}.
However there were certainly some manifest peaks corresponding to recurrent fragments, especially high for some patients, most frequently those diagnosed with cancer. It is also apparent that the individual absorbance profiles are measured at different mean levels for different patients and with different variability. A simple logarithmic transformation and recentering w.r.t. the individual mean were performed providing some more stable evidence of differential profiles among groups. However there are some specific profiles which seem pretty much overlapped,
among different subgroups 
although with some different overall pattern (Figure~\ref{LogCen}).

\begin{figure}
\begin{center}
\includegraphics[height=15cm, width=17cm]{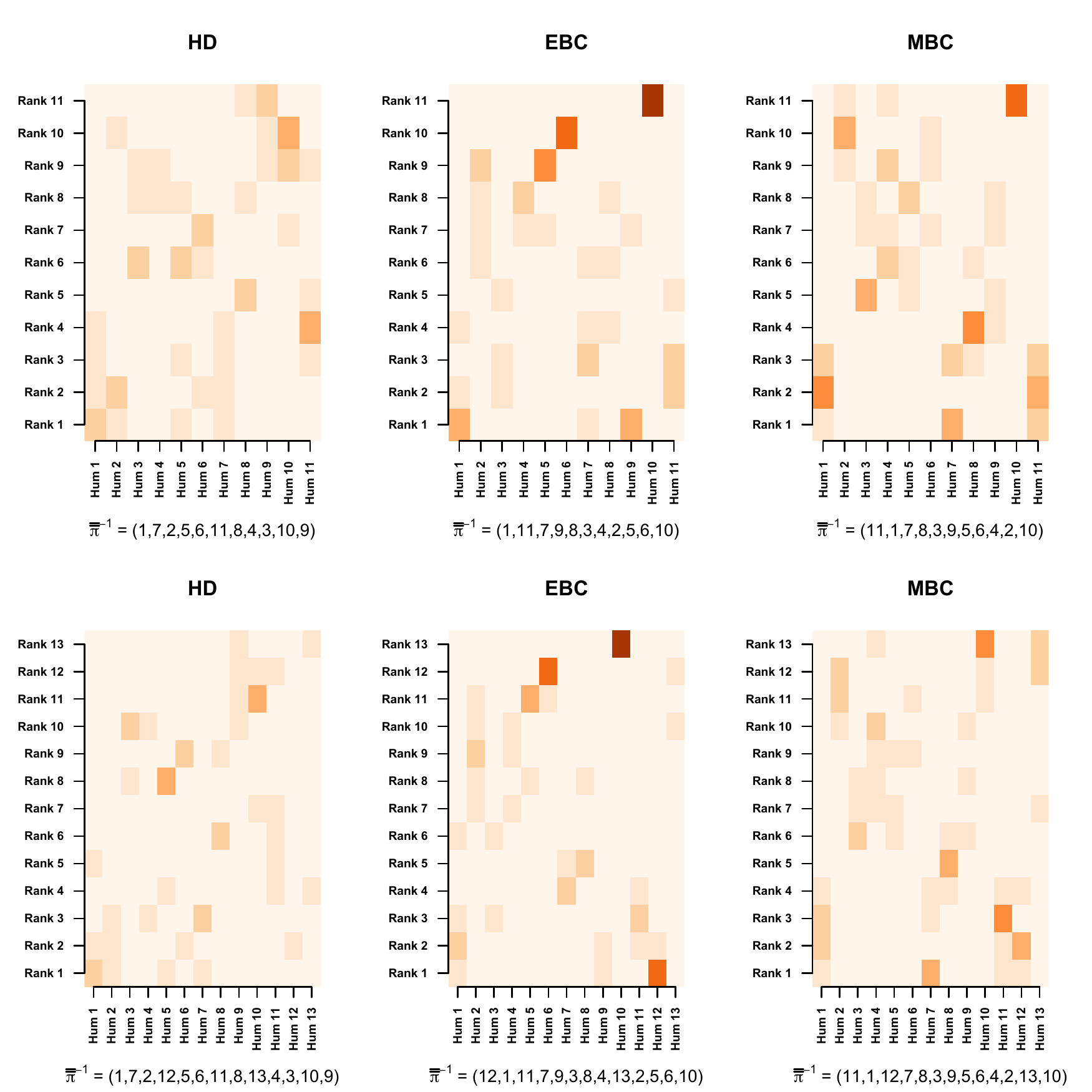} 
\caption
{Image plots of the first-order marginal matrices for the three groups of patients in the LFPD study:
HD  $=$ healthy (left),
EBC $=$ diagnosed with early stage breast cancer (center),
MBC $=$ diagnosed with metastatic breast cancer (right).
Upper panel refers to the data with $K=11$ protein fragments whereas the lower one concerns the $K=13$ case with the addition of Hum 12 and Hum 13, indicating respectively the whole HER2 oncoprotein (positive control) and the empty phage vector (negative control). The Borda ordering $\overline{\overline{\pi}}^{-1}$ lists items taken in order from the highest to the lowest mean rank.}
\label{f:figure4}
\end{center}
\end{figure}

Since data emerged from the development of an innovative technology, miscalibrations or 
inaccuracies of the measuring device 
may occur 
and/or 
subject-specific characteristics may 
alter somehow
the observed numerical outcome, 
making it more difficult
to adjust  
the statistical analysis based
on raw or ad-hoc pre-processed
data.
 Unfortunately, for this kind of data 
a consolidated and fully-shared normalization technique
 is lacking.
For all these reasons, rather than basing our analysis 
on the quantitative output of the LFPD study, we verified the possible usefulness of
the ranking profiles 
as a more robust and
unambiguously-defined evidence,
capable to capture and characterize the sample heterogeneity 
w.r.t. the disease status
and specific characteristic profiles of each subgroup. 
Hence, we first derived ordered sequences ranking the absorbance levels of the individual protein fragments taken in decreasing order (Rank 1=highest value, Rank $K$=lowest value). We
performed a simple exploratory analysis by cancer status 
computing both the 
$K\times K$ first-order 
marginal 
matrices $\hat{M}$, 
where the generic entry $\hat{M}_{ij}$ indicates the observed relative frequency that item $i$ is ranked $j$-th, and the so-called Borda orderings $\overline{\overline{\pi}}^{-1}$, listing items taken in order from the highest to the lowest mean rank. 
These matrices are displayed 
as image plots in Figure~\ref{f:figure4},
%
%
together with the
Borda sequences
on the
bottom of each panel. 
The color intensity is an increasing 
function of the corresponding observed frequency. 
The analysis of the first-order marginals matrices suggests that very often some protein fragments are associated with lower ranks, as pointed out by the presence of darker rectangles in correspondence of bottom positions.
This constantly occurs for all disease subgroups with Hum 10 but some interesting differential evidence emerges from EBC subjects with Hum 5 and 6, from MBC with Hum 2 and 13 and also from HD patients with Hum 9 (Figure~\ref{f:figure4}).
Such a precious discriminant information could be better captured by our EPL. 
To validate this claim 
we fitted both the PL and the new EPL to the three disease subgroups separately. For the former we used known orders, alternatively forward (PL-$\rho_{1}$) and backward (PL-$\rho_{2}$), whereas for the latter the reference order is a parameter to be estimated.
Estimation performances are shown in terms of BIC values in Table~\ref{t:bicPLnew} and reveal that
the fit of the EPL is better or at most comparable with those relative to the PL with fixed reference orders. The interest in relaxing the canonical forward assumption is supported also by the BIC values for the PL-$\rho_{2}$, showing such a model constantly outperforms the PL-$\rho_{1}$ when fitted to HD and MBC subjects. These BIC results represent a strong evidence motivating the need of an extension of the PL.
%
\begin{table}[h]
\caption{BIC values resulting from the MLE of the PL ($\rho_1$ $=$ forward and $\rho_2$ $=$ backward reference order) and of the EPL on the three disease groups for a different number $K$ of binding probes included in the ranking. Groups of patients are defined as follows: HD  $=$ healthy,
EBC $=$ diagnosed with early stage breast cancer, and
MBC $=$ diagnosed with metastatic breast cancer.}
\renewcommand{\arraystretch}{1.2} 
\label{t:bicPLnew}
\begin{center}
\begin{tabular}
 {l|*2{|*{3}{>{$}c<{$}}}}

& \multicolumn{3}{c|}{$K=11$}
& \multicolumn{3}{c }{$K=13$} \\
\cline{2-7}
Model\quad$ $
& \text{HD} & \text{EBC} & \text{MBC}
& \text{HD} & \text{EBC} & \text{MBC} \\
\hline
PL-$\rho_1$   & 694.04          & 776.13          & 499.46          & 899.63          & \mathbf{1025.71} & 658.44          \\
PL-$\rho_2$   & 685.85          & 804.61          & 498.67          & 894.44          & 1039.45          & 652.15 \\
EPL           & \mathbf{676.93} & \mathbf{773.17} & \mathbf{473.90} & \mathbf{873.71} & 1026.61          & \mathbf{630.05} \\
\end{tabular}
\end{center}
\end{table}
Subsequently we considered a more comprehensive analysis in a mixture model setting. With this approach we aimed at:
\begin{itemize}
\item addressing the heterogenous nature of the LFPD data using the evidence provided by the orderings of absorbance levels;
\item assessing if and how the path in the sequential ranking process 
can have an impact on the final model-based classification 
of experimental units and 
select the most appropriate one.
\item looking for possible characteristic subgroups related to the disease status;
\item characterizing each subgroup with the estimates of the cluster-specific parameters.
\end{itemize}

\subsection{Empirical findings from mixture models fitted to LFPD data}

Considering all 67 available orderings 
we fitted mixtures of DB with $d=d_K$ (DBmix), mixtures of PL with both forward and backward reference order (PLmix-$\rho_1$ and PLmix-$\rho_2$) and mixtures of EPL (EPLmix), the novel model we presented in section~\ref{ss:proposal} where $\rho$ is a parameter to be inferred. All mixtures have been implemented with a number of components varying from $G=1$ to $G=7$. Of course, the case $G=1$ coincides with the assumption that observations come from a homogeneous population without an underlying group structure.
We separately applied the mixture models to the ranking of absorbance levels relative to the $K=11$ partially overlapping protein fragments as well as to the $K=11+2$ binding probes (spots), when we additionally included the whole HER2 oncoprotein (positive control) and the empty phage vector (negative control). 

Focusing on the BIC for $G=1$ compared to $G>1$, the MLE of the DBmix provided an overall 
evidence in favor of a
heterogeneity
when 
both
$K=11$ or $K=13$ binding probes are considered.
We highlighted a remarkable decreasing behavior for the associated BIC,
which persists when fitting is carried out up to $G=10$ components
as shown in 
Table~\ref{t:bicDB}.
\begin{table}[]
\caption{BIC values resulting from the MLE of the DBmix on the LFPD data with a varying number $G$ of components, when either $K=11$ or $K=13$ binding probes are included in the ranking.}
\renewcommand{\arraystretch}{1.2} 
\label{t:bicDB}
\begin{center}
\begin{tabular}{lcccccccccc}
& \multicolumn{10}{c}{$G$} \\
\cline{2-11}
& 1 & 2 & 3 & 4 & 5 & 6 & 7 & 8 & 9 & 10 \\
\hline
$K=11$
& 2078.77 & 2003.65 & 1940.86 & 1899.33 & 1882.32 & 1863.17 & 1846.98 & 1829.81 & 1817.06 & 1798.12\\
$K=13$
& 2700.02 & 2617.66 & 2551.38 & 2512.19 & 2483.25 & 2451.38 & 2421.71 & 2392.10 & 2366.78 & 2342.60\\
\hline
\end{tabular}
\end{center}
\end{table}
\begin{center}
\begin{figure}[b]
\centering
\subfloat{
\includegraphics[width=0.4\textwidth]{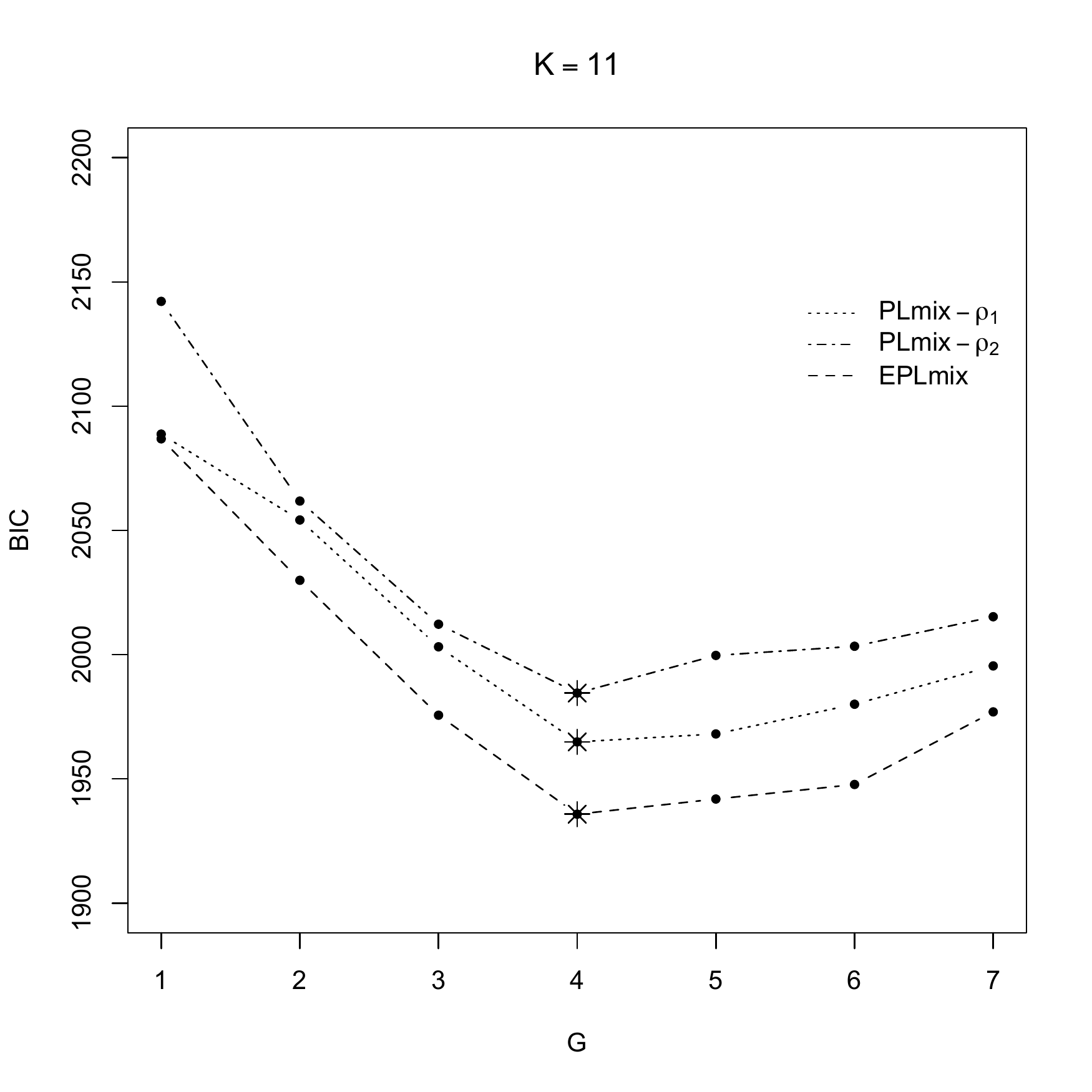}
\label{fig:subfig2}}
\qquad
\subfloat{
\includegraphics[width=0.4\textwidth]{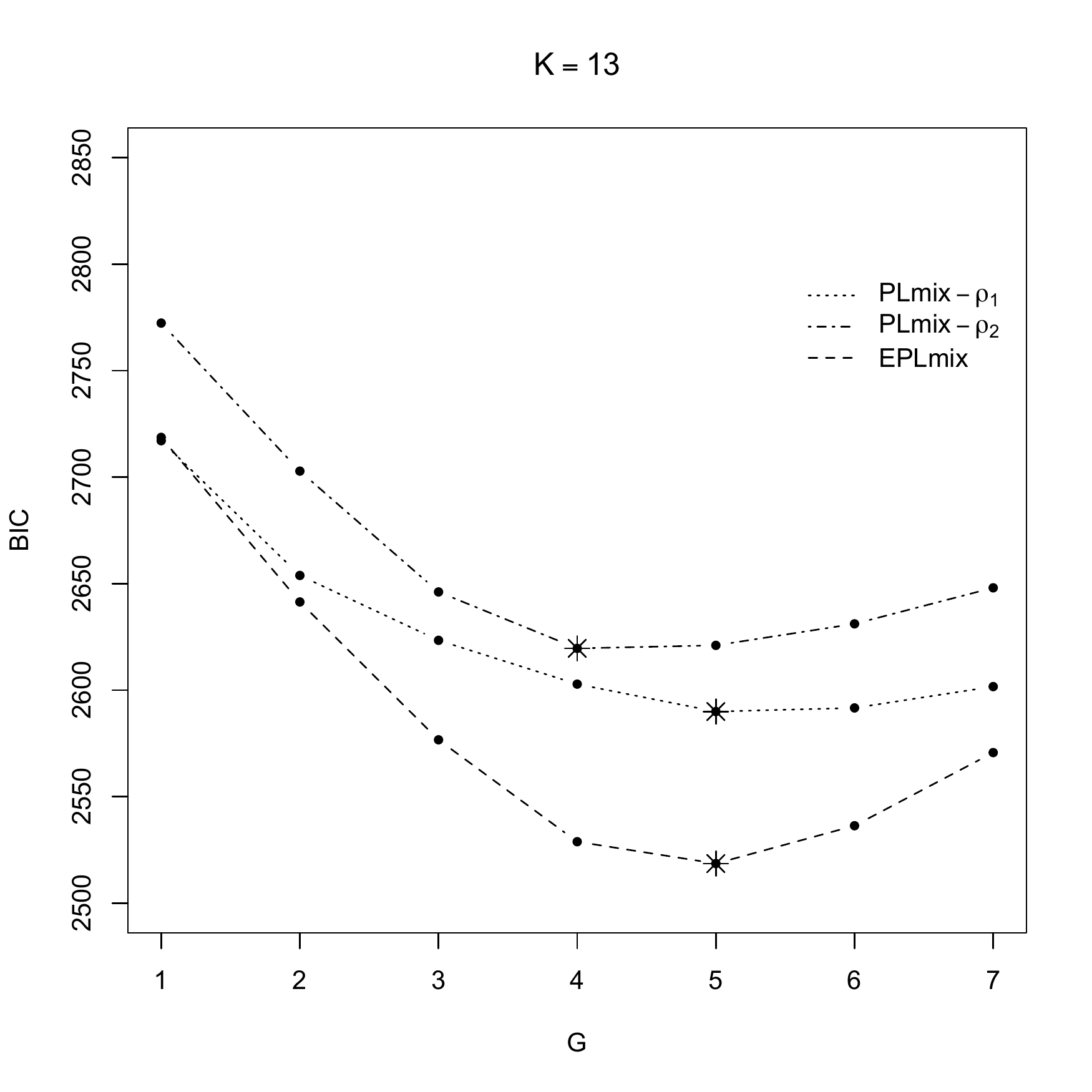}
\label{fig:subfig4}}
\caption{BIC trends resulting from the MLE of the PLmix-$\rho_1$, the PLmix-$\rho_2$ and the EPLmix on the LFPD data with a varying number $G$ of mixture components, when either $K=11$ (left) or $K=13$ (right) binding probes are included in the ranking. The symbol \ding{87} indicates the minimum BIC values for the final selection of the number of groups.}
\label{fig:BIC}
\end{figure}
\end{center}
Indeed, fitting DBmix with an increasing number of groups pointed out a particular feature of the DB, probably due to the sparse nature of LFPD data. We remind, in fact, that in the present application the sample size is small w.r.t. to the cardinality ($11!$ or $13!$) of the discrete ranking space. As the value of $G$ in the DBmix increases, some components start to represent 
just a 
single observation. This can be explained, perhaps, by the fact that, once the modal ranking $\sigma$ has been fixed, DB has only one remaining parameter fitting the shape of the uncertainty. It follows that for these components the concentration parameter $\lambda$ is typically estimated as a very high value. This behavior, of course, could make the DBmix model not sufficiently parsimonious and promising in some sparse-data situations, because its fit would lead to a more 
sparse
clustering of the observations and 
to a less 
enlightening
inferential findings.
%
When
stage-wise models have been
fitted to 
the LFPD data,
we found 
again evidence in favor of the heterogeneous
structure,
as indicated in Table~\ref{t:bicEPLnew}; in the case $K=11$ all types of mixtures consistently identify 4 groups in the sample, whereas one additional component is selected by both PLmix-$\rho_1$ and EPLmix when also the control probes are included in the ordered sequences.
%
Bold BIC values in Table~\ref{t:bicEPLnew} indicate the EPLmix as the best model and are, in both cases, significantly smaller than those of the competing mixtures. Indeed, this is constantly true for every considered dimension $G$ of the mixture, as elucidated by Figure~\ref{fig:BIC}. This proves the successful introduction of the discrete parameter $\rho$ which drastically improves the fitting of the data. Moreover, the EPLmix exhibits a good accuracy in the discrimination of sample units w.r.t. the real disease status.  
\begin{table}[]
\caption{BIC values and number $G$ of components of the best PLmix-$\rho_1$, PLmix-$\rho_2$ and EPLmix fitted to LFPD data for different number $K$ of binding probes included in the ranking.}
\renewcommand{\arraystretch}{1.2} 
\label{t:bicEPLnew}
\begin{center}
\begin{tabular}
 {l|*2{|*{2}{>{$}c<{$}}}}
Mixture\quad$ $
& \multicolumn{2}{c|}{$K=11$}
& \multicolumn{2}{c }{$K=13$} \\
\cline{2-5}
Model\quad$ $
& \text{BIC} & \text{$G$}
& \text{BIC} & \text{$G$} \\
\hline
PLmix-$\rho_1$     & 1964.87          & 4 & 2589.93          & 5 \\
PLmix-$\rho_2$     & 1984.53          & 4 & 2619.60         & 4 \\
EPLmix             & \mathbf{1935.77} & 4 & \mathbf{2518.56} & 5 \\
\end{tabular}
\end{center}
\end{table}
%
The two resulting clusterings agree with the most relevant distinction of the real disease status (healthy/non healthy), as pointed out in Tables~\ref{t:agreeEPL}\subref{t:agree1new} and~\ref{t:agreeEPL}\subref{t:agree2new}. Specifically, collapsing also the model-based group membership into this basic bipartition
we recognize that healthy subjects are well isolated with only 1 or 2 false positive cases, whereas for diseased patients we have $7$ misclassifications when $K=11$ but only 2 with the addition of the control spots, see Tables~\ref{t:agreeEPL}\subref{t:agree1new} and~\ref{t:agreeEPL}\subref{t:agree2new}. As expected, the inclusion of the positive and negative controls produced a fruitful discriminant evidence, suggested by the global reduction of misclassifications for clusterings based on 13 ranks. Healthy patients are always modeled with two components in all mixtures fitted to the LFPD data. This
hints at possibly different subtypes of healthy profiles. In fact, we easily verified that such subdivision reflects two different absorbance patterns in cancer-free units, made evident in Figure~\ref{Raw} by the green broken lines: one subgroup whose immune defenses essentially did not react at all to the exposition with the HER2 oncoprotein (lower panel) and those with some manifest and characterized binding profile (upper panel).   
\begin{table}[b]
\caption{Correspondence between the model-based clustering derived by the MLE of the EPLmix and the true disease status of the LFPD experimental units:
HD  $=$ healthy,
EBC $=$ diagnosed with early stage breast cancer and
MBC $=$ diagnosed with metastatic breast cancer.}
\label{t:agreeEPL}
\centering
\subfloat[$K=11$]  
{\begin{tabular}{lcccc}
 & \multicolumn{4}{c}{Group} \\
\cline{2-5}
 \multicolumn{1}{c}{Disease Status} & 1 & 2 & 3 & 4\\ 
  \hline
HD  &  0  & 2  & 10 &  8 \\ 
EBC &  13 & 12 & 2  &  1 \\ 
MBC &  0  & 15 & 3  &  1 \\ 
   \hline
\end{tabular}
\label{t:agree1new}}\quad
\subfloat[$K=13$]
{\begin{tabular}{lccccc}
& \multicolumn{5}{c}{Group} \\
\cline{2-6}
 \multicolumn{1}{c}{Disease Status} & 1 & 2 & 3 & 4 & 5 \\ 
  \hline
HD  &   1 &  10 & 0 & 1 & 8 \\ 
EBC &  12 &  0  & 9 & 7 & 0 \\ 
MBC &  14 &  2  & 0 & 3 & 0 \\ 
   \hline
\end{tabular}
\label{t:agree2new}}
\end{table}
%
On the other hand, among the selected components which can be categorized as corresponding to diseased patients, the sub-classification between EBC and MBC is only partially recovered, especially for the latter group of patients. This is proved by the presence of at least one model-based group entirely composed of EBC subjects in all fitted mixtures, whereas MBC always belong to mixed-type components.

\begin{figure}[t]
\centering
\subfloat{ 
\includegraphics[width=0.33\textwidth]{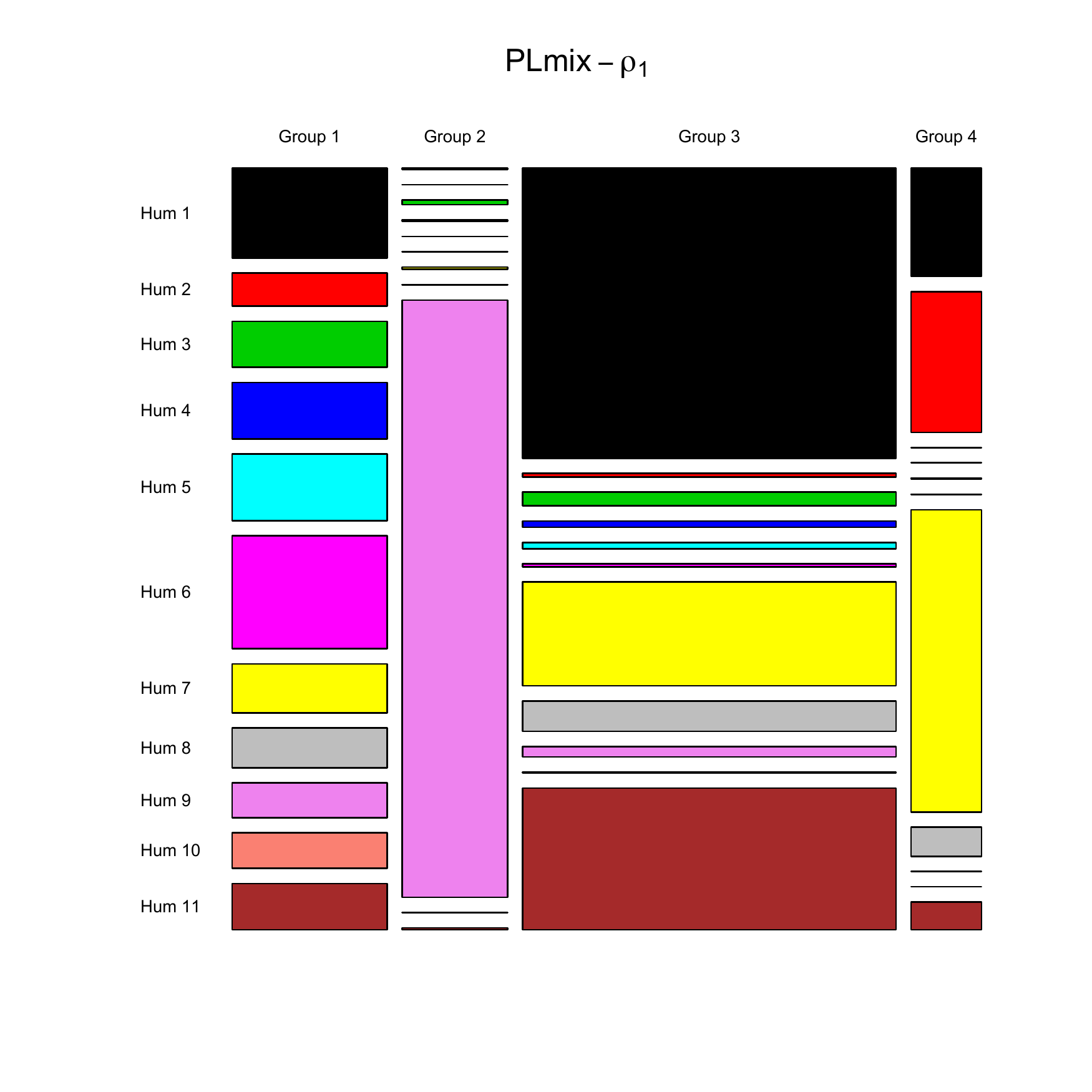}  
\label{fig:SuppEst11_forw}}
\subfloat{ 
\includegraphics[width=0.33\textwidth]{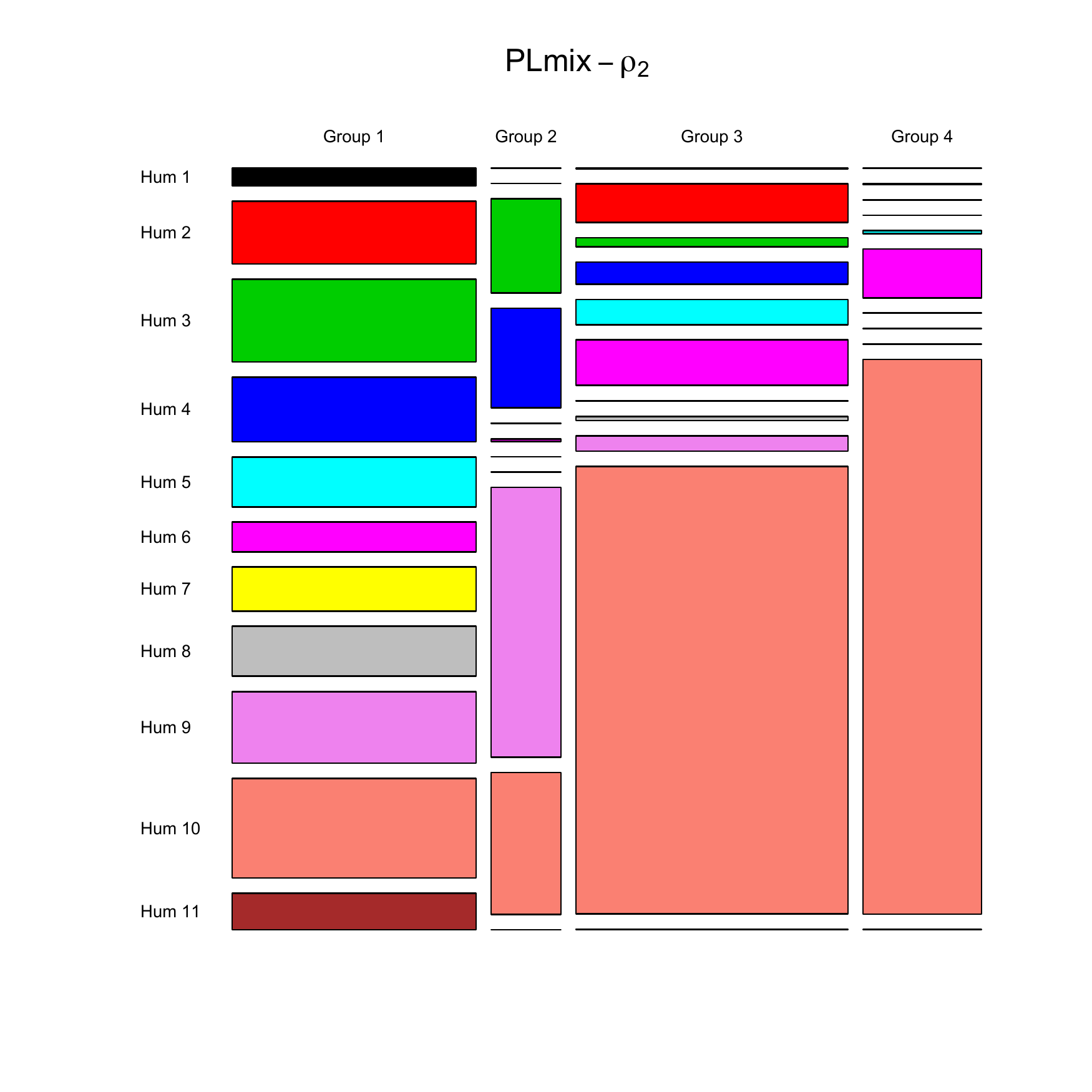}
\label{fig:SuppEst11_back}}
\subfloat{ 
\includegraphics[width=0.33\textwidth]{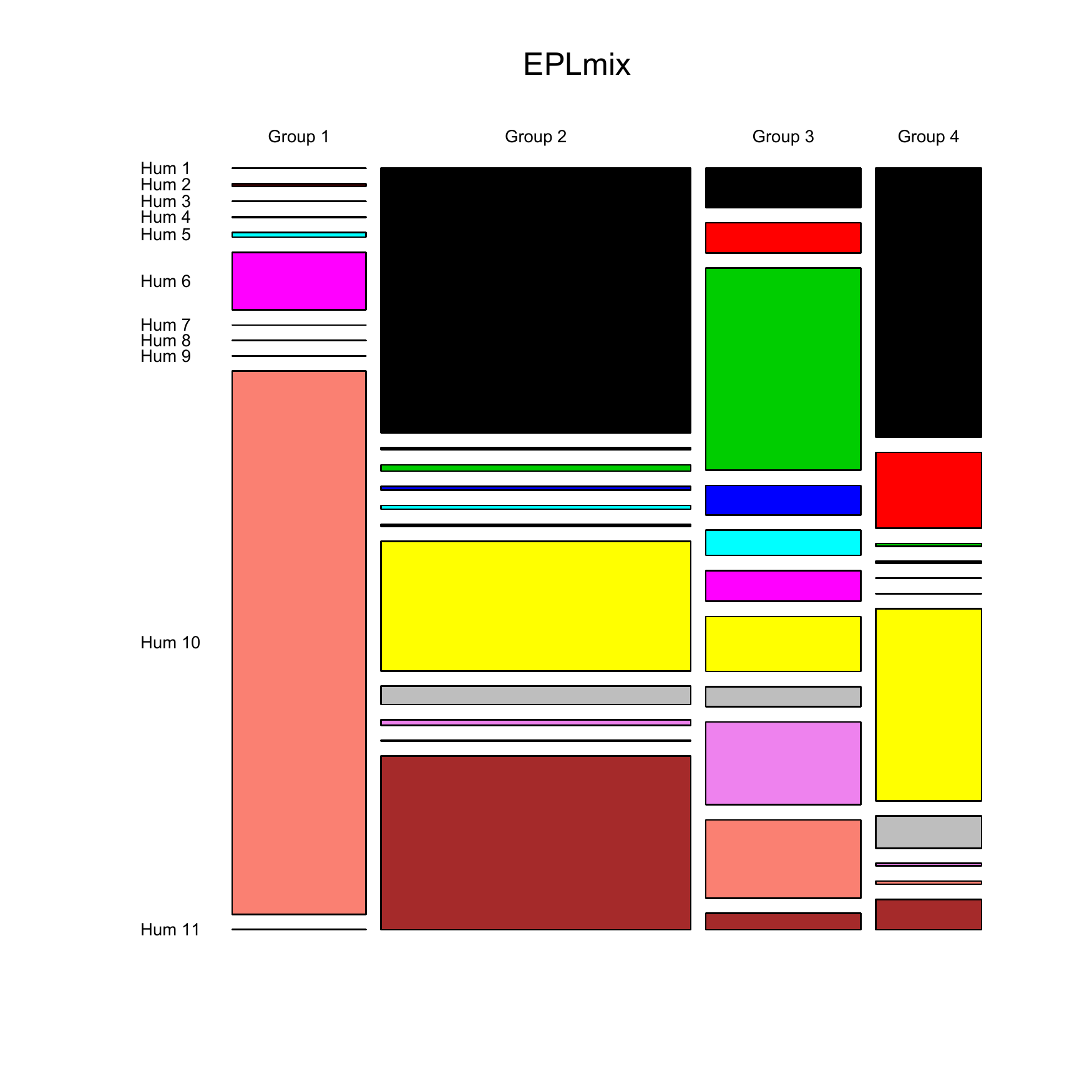}
\label{fig:SuppEst11}}
\qquad
\subfloat{ 
\includegraphics[width=0.33\textwidth]{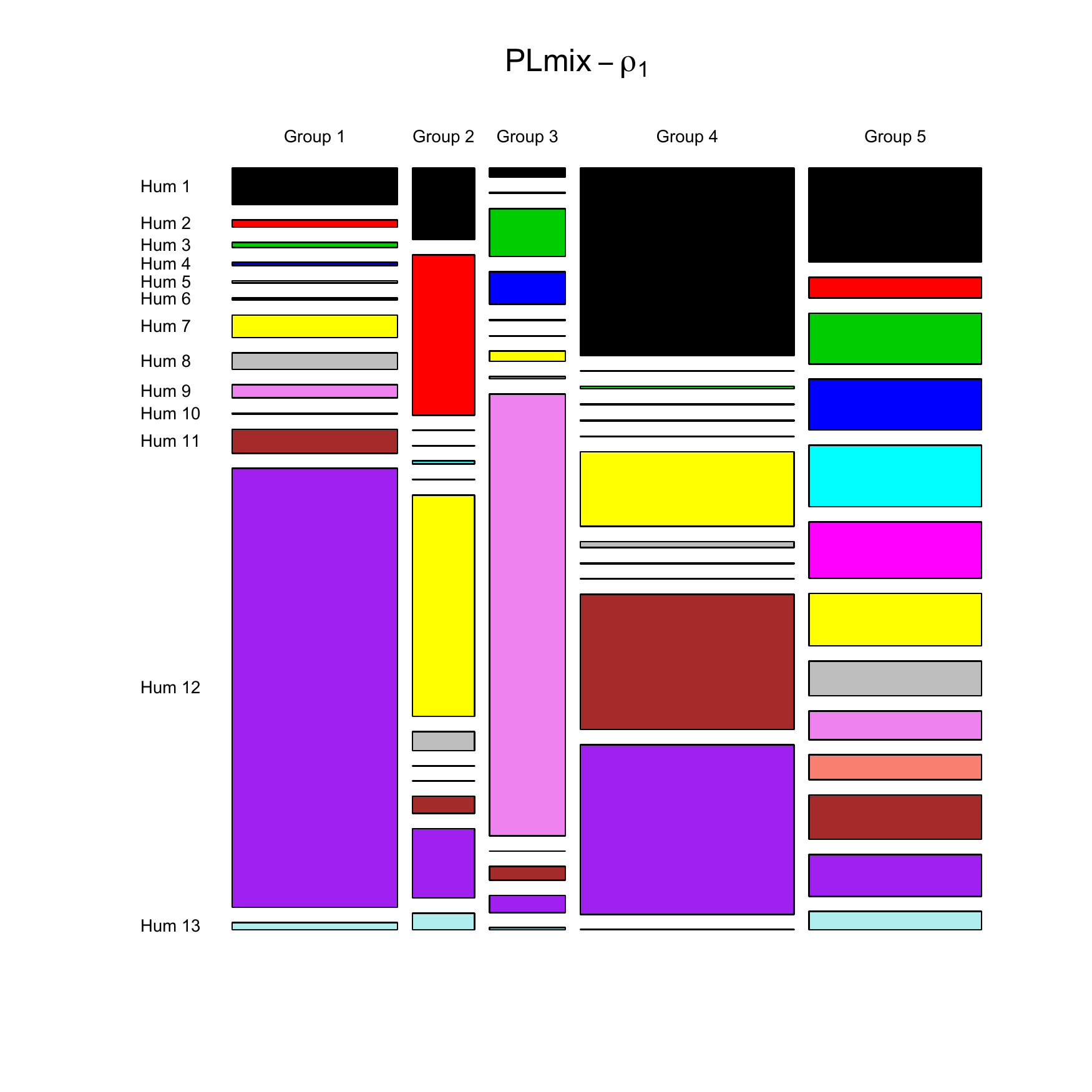}
\label{fig:SuppEst13_forw}}
\subfloat{ 
\includegraphics[width=0.33\textwidth]{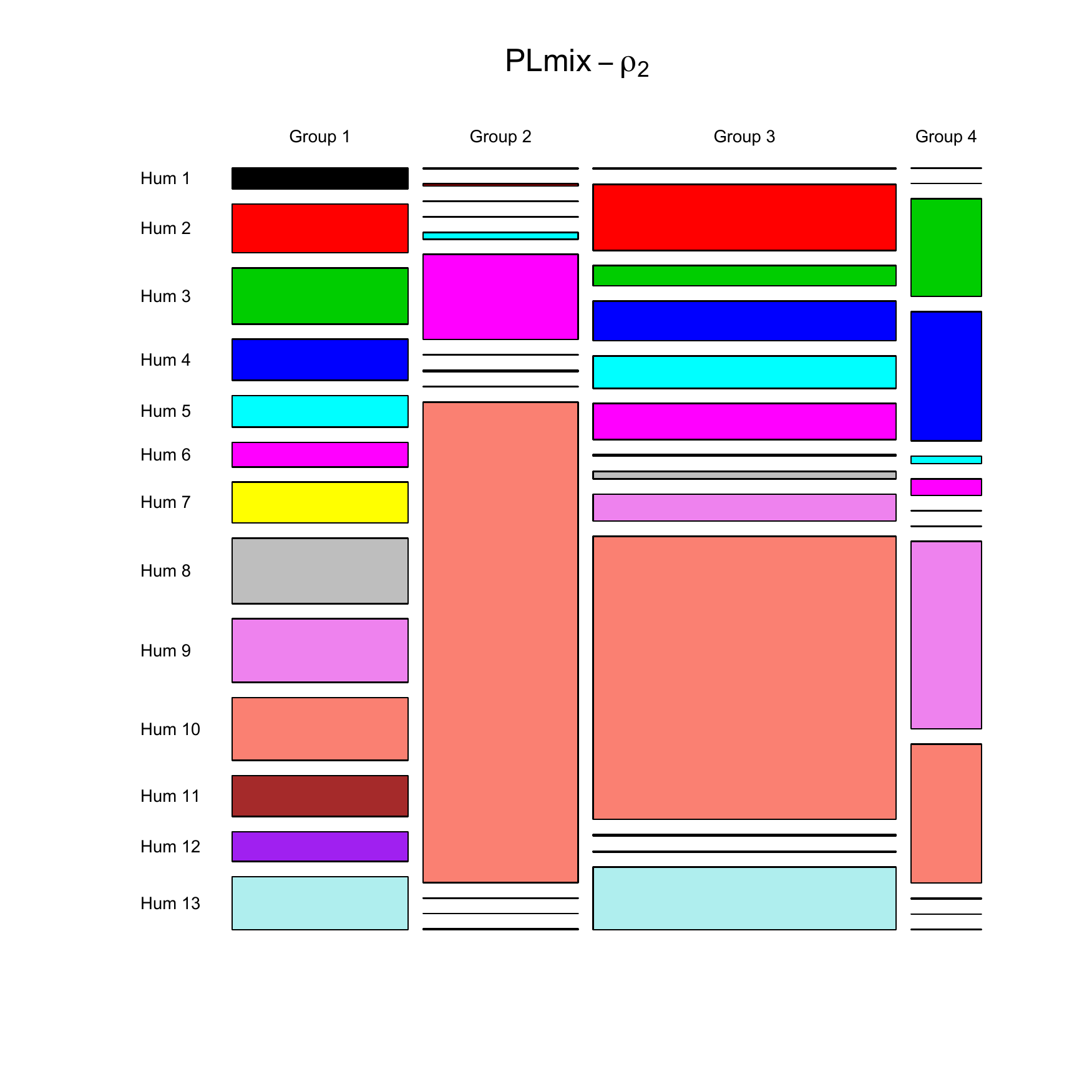}
\label{fig:SuppEst13_back}}
\subfloat{ 
\includegraphics[width=0.33\textwidth]{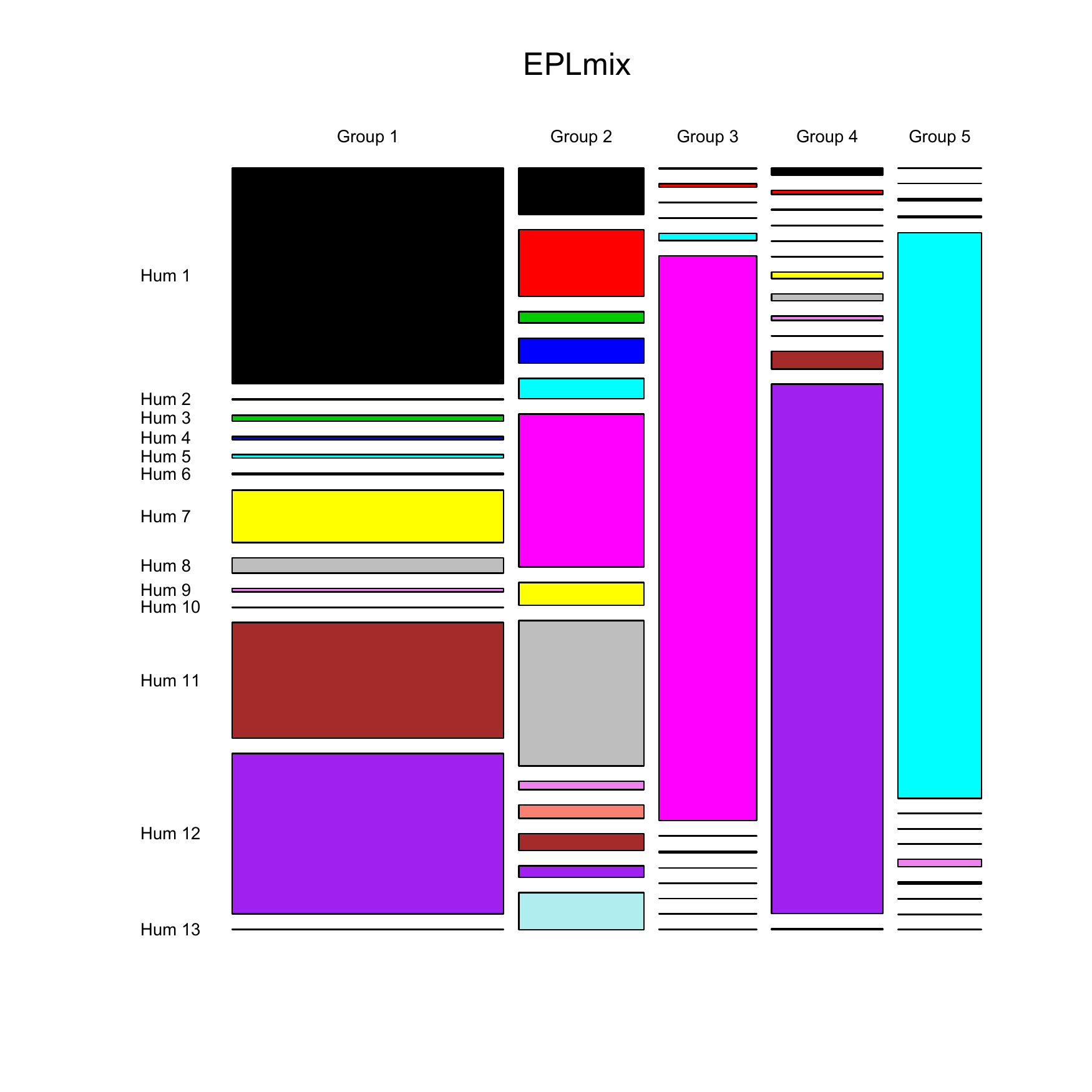}
\label{fig:SuppEst13}}
\caption{Support parameter estimates represented via mosaic plots for the best PLmix-$\rho_1$, PLmix-$\rho_2$ and EPLmix fitted to the LFPD data. Bar widths are proportional to the group weights. Upper panel refers to the data with $K=11$ protein fragments whereas the lower one concerns the $K=13$ case with the addition of Hum 12 and Hum 13, indicating respectively the whole HER2 oncoprotein (positive control) and the empty phage vector (negative control).}
\label{fig:SuppEst}
\end{figure}

The varying correspondence between the real cancer status and the inferred clustering 
structure 
confirms the presumed dependence of the classification results on the adopted reference ranking process $\rho$. 
Furthermore, the good agreement obtained with the EPLmix, and pointed out by Tables~\ref{t:agreeEPL}\subref{t:agree1new} and~\ref{t:agreeEPL}\subref{t:agree2new}, suggests that researchers 
should not focus exclusively on differential epitope identification 
but should extend their analysis 
considering also more general global understanding of differential bindings. 
Hence, in order to characterize disease groups w.r.t. ranking profiles it is interesting to interpret the component-specific
modal orderings (Table~\ref{t:sumestNew1}),
derived by ordering the corresponding support parameter estimates (Figure~\ref{fig:SuppEst}).
Weights and reference order estimates for the identified clusters are shown in Table~\ref{t:sumestNew2}. 
\begin{table}[t]
\footnotesize 
\addtolength{\tabcolsep}{-3pt}    
\renewcommand{\arraystretch}{1.2} 
\caption{Modal orderings and composition w.r.t. the real cancer status of the components identified with the best PLmix-$\rho_1$, PLmix-$\rho_2$ and EPLmix fitted to LFPD data, for a different number $K$ of binding probes included in the ranking. ``D.C.'' stands for ``disease composition'' listing sequentially the number of
HD  $=$ healthy,
EBC $=$ diagnosed with early stage breast cancer and
MBC $=$ diagnosed with metastatic breast cancer
patients in each group.}
\resizebox{15cm}{!}{
\begin{minipage}{\textwidth}
\begin{center}
\begin{threeparttable}
\begin{tabular}
 {l|*2{|*{3}{>{$}c<{$}}}}
Mixture\quad$ $
& \multicolumn{3}{c|}{$K=11$}
& \multicolumn{3}{c }{$K=13$} \\
model\quad$ $
& g & \text{D.C.} & \hat\sigma_g^{-1}
& g & \text{D.C.} & \hat\sigma_g^{-1} \\
\hline
PLmix-$\rho_1$
& 1 & (11,1,3)  & (6,1,5,4,7,11,3,8,10,9,2)$*$    & 1 & (2,11,3)   & (12,1,11,7,8,9,2,13,3,4,6,5,10) \\
& 2 & (0,10,0)  & (9,3,7,11,4,1,8,2,5,6,10)       & 2 & (6,0,0)    & (7,2,1,12,8,11,13,5,6,10,4,3,9) \\
& 3 & (3,17,15) & (1,11,7,8,3,9,4,5,2,6,10)       & 3 & (0,7,0)    & (9,3,4,12,11,7,1,8,13,2,5,6,10) \\
& 4 & (6,0,1)   & (7,2,1,8,11,5,6,4,10,3,9)       & 4 & (0,7,14)   & (1,12,11,7,8,3,5,9,4,6,2,13,10) \\ 
&   &           &                                 & 5 & (12,3,2)   & (1,5,6,7,3,4,11,12,8,9,10,2,13)$*$ \\
\hline                
PLmix-$\rho_2$
& 1  & (14,5,3)   & (1,6,11,7,5,8,2,4,9,3,10)$*$  & 1 & (12,2,2)  & (1,6,12,5,7,11,4,2,13,3,10,9,8)$*$ \\
& 2  & (6,0,1)    & (7,2,1,8,11,5,6,3,4,10,9)     & 2 & (0,15,0)  & (12,9,3,7,4,11,1,13,8,2,5,6,10) \\
& 3  & (0,12,15)  & (1,7,11,8,3,9,4,5,2,6,10)     & 3 & (1,11,17) & (12,1,11,7,8,3,9,5,6,4,13,2,10) \\
& 4  & (0,11,0)   & (9,3,4,7,11,1,8,2,5,6,10)     & 4 & (7,0,0)   & (7,2,1,12,8,13,11,5,6,3,4,10,9) \\
\hline
EPLmix
& 1 & (0,13,0)  & (9,8,1,3,11,7,2,4,5,6,10)    & 1 & (1,12,14)  & (12,1,11,7,8,3,4,9,5,6,2,13,10) \\
& 2 & (2,12,15) & (1,11,7,8,9,3,4,5,6,2,10)    & 2 & (10,0,2)   & (5,2,11,4,3,6,10,7,8,9,12,1,13) \\
& 3 & (10,2,3)  & (5,4,11,1,6,3,10,2,9,7,8)    & 3 & (0,9,0)    & (9,12,11,3,1,4,7,2,13,8,5,6,10) \\
& 4 & (8,1,1)   & (7,2,1,8,11,5,6,4,10,9,3)    & 4 & (1,7,3)    & (12,9,1,11,13,3,8,7,2,4,5,6,10) \\
&   &           &                              & 5 & (8,0,0)    & (11,2,1,6,12,8,13,5,7,10,4,3,9) \\
\end{tabular}
\begin{tablenotes}[para,flushleft]
Note: The symbol $*$ indicates mixture components which are very close to the UM.
\end{tablenotes}
  \end{threeparttable}
\end{center}
\end{minipage}
}
\label{t:sumestNew1}
\end{table}
\begin{table}[b]
\footnotesize 
\addtolength{\tabcolsep}{-3pt}    
\renewcommand{\arraystretch}{1.2} 
\caption{Mixture weights and reference order estimates of the best PLmix-$\rho_1$, PLmix-$\rho_2$ and EPLmix fitted to LFPD data, for a different number $K$ of binding probes included in the ranking.}
\centering
\resizebox{15cm}{!}{
\begin{minipage}{\textwidth}
\begin{center}
\begin{tabular}
 {l|*2{|*{3}{>{$}c<{$}}}}
Mixture\quad$ $
& \multicolumn{3}{c|}{$K=11$}
& \multicolumn{3}{c }{$K=13$} \\
model\quad$ $
& g & \hat\omega_g & \hat\rho_g
& g & \hat\omega_g & \hat\rho_g \\
\hline
PLmix-$\rho_1$
& 1 & .22  & \rho_1    & 1 & .24   & \rho_1 \\
& 2 & .15  & \rho_1    & 2 & .09   & \rho_1 \\
& 3 & .53  & \rho_1    & 3 & .11   & \rho_1 \\
& 4 & .10  & \rho_1    & 4 & .31   & \rho_1 \\ 
&   &           &      & 5 & .25   & \rho_1 \\
\hline
PLmix-$\rho_2$
& 1 & .35  & \rho_2    & 1 & .25  & \rho_2 \\
& 2 & .10  & \rho_2    & 2 & .22  & \rho_2 \\
& 3 & .39  & \rho_2    & 3 & .43  & \rho_2 \\
& 4 & .16  & \rho_2    & 4 & .10  & \rho_2 \\
\hline
EPLmix
& 1 & .19  & (11,10,9,7,8,4,2,3,6,5,1)    & 1 & .39    & (2,1,3,4,5,6,8,9,7,10,11,12,13) \\
& 2 & .44  & (1,2,3,4,6,5,7,8,9,10,11)    & 2 & .18    & (6,9,2,12,13,4,8,1,3,7,11,5,10) \\
& 3 & .22  & (6,9,7,10,4,5,8,2,1,11,3)    & 3 & .14    & (12,11,8,10,9,5,7,6,3,4,2,1,13) \\
& 4 & .15  & (3,1,2,4,5,9,11,10,8,7,6)    & 4 & .17    & (1,4,3,7,8,2,9,5,6,10,12,13,11) \\
&   &      &                              & 5 & .12    & (8,13,12,10,11,1,6,7,4,5,9,2,3) \\
\end{tabular}
\end{center}
\end{minipage}}
\label{t:sumestNew2}
\end{table}
%
%
%
%
%
%
Focusing on the 
analysis 
based on $13$ binding probes,
we stress that in all 
the best fitted models
the positive control probe, labeled as Hum 12, recurrently occupies top positions in modal orderings of EBC and EBC+MBC mixture components. We remind that Hum 12 denotes the absorbance level 
corresponding 
to the entire HER2 oncoprotein. Thus, in theory, its level should reflect 
the total binding and it is reasonably expected to be higher than absorbance 
level detected in limited portions of the oncoprotein. 
On the other hand, immunological response in healthy patients may either be unaffected by the exposition with the HER2 oncoprotein or yield a mild binding. This means an exchangeability of binding probes in the ordering of absorbance levels, which is typical under the UM. These aspects reinforce the presence of Hum 12 in top positions as a signal that the immunological response actually occurred and hence it can be interpreted as a distinguishing feature of the unhealthy patients. It turns out that with our wildly fluctuating LFPD data it is not possible to identify a simple threshold for the raw (or normalized) binding outcome to discriminate unhealthy patients. This is better achieved using binding profiles based on rankings.  
Moreover, 
the combination of 
Hum 12 with the pattern (Hum 1, Hum 11, Hum 7) in top positions seems to characterize mixed (EBC+MBC) 
diseased groups, such as the first and the fourth components in PLmix-$\rho_1$, 
the third one in PLmix-$\rho_2$ and the first one in EPLmix. 
In fact the protein fragments Hum 1, Hum 11 and 7 were already recognized 
in~\citep{Gabrielli:al} as the relevant epitopes. 
In EBC-specific components, similar results are valid for the fragment pair (Hum 9,  Hum 3) which, together with the positive control, occupies 
the very first top positions, see for example the third group in PLmix-$\rho_1$ and the second one in PLmix-$\rho_2$; 
this means that for some EBC patients the binding reaction mainly occurs in a different section of the oncoprotein, improving the discrimination of this subgroup 
among diseased patients.
Relevant findings can be also highlighted for healthy patients. The absent or negligible immunological response observed for some of them is well described by estimated models with 
a component which is very close to the UM, as shown by the corresponding inferred value $\underline{\hat p}_g$. In this case the modal orderings are poorly representative, so we marked them with the symbol $*$ in Table~\ref{t:sumestNew1}.
These UM components involve prevalently HD patients. They
are also included in another more characterized mixture component.
 The interpretation of the non-uniform component parameters suggests that some HD subjects share the epitopes Hum 1 and Hum 7 with other patients but they also have a distinctive Hum 2 in top positions; Hum 11, instead, appears in middle positions. 
We can also look at 
low absorbance patterns, if bottom ranks can be regarded as 
meaningful 
{\em signatures}
for the problem at hand. 
Note, for example, that whereas Hum 10 appears consistently in last positions for almost all of the fitted components, Hum 9 seems to be a sort of ``anti''-epitope 
signature 
for HD units; the same role is played by the Hum pattern (Hum 5, Hum 6) for EBC subjects. Another interesting feature regards Hum 13; it corresponds to the empty phage vector hence, theoretically, one would expect it to be associated with bottom ranks whereas this 
is true specifically for those groups
composed for the most part of MBC units, see for example the fourth component in the PLmix-$\rho_1$, the third one in the PLmix-$\rho_2$ and the first one in the EPLmix. Hence, a minimum absorbance level in Hum 13 could be an important indication to discriminate MBC patients, the subgroup which is only weakly characterized by the present analysis. Similar observations are valid for the case $K=11$ omitting, naturally, Hum 12 and Hum 13. 
Finally, we 
remark that the model selected as the best in terms of the BIC is the EPLmix, which involves $69$ parameters in the case of $K=13$. 

\subsection{Alternative quantitative data analysis}



Now we show that our analysis based on ranked data
and EPL mixture model 
compares favorably with a more conventional approach relying 
on quantitative data.  
We implemented the 
flexible mixture of multivariate normal distributions (MNorm-mix) 
with the R package \texttt{mclust} described in~\citep{Fraley:Raftery}. 

As urged in section~\ref{LFPDana}, we must preliminarily
decide 
whether there exists a more 
appropriate way of transforming and rescaling the original quantitative measures. Because a consolidated normalization method is lacking for this type of experiments, we worked with 3 alternative reasonable options:
original raw data, the log-transformed absorbances and the rescaled log-transformed absorbances so that the individual average log-absorbance of all the considered spots is null for each patient. 
Results derived from the quantitative analysis are very different according to 
which measurement scale is used in the input data. 
In fact, only with the raw data the best fitting mixture model 
provides evidence in favor of an heterogeneous model, 
namely a mixture with 
$G=3$ components. 
However, as shown in Table~\ref{t:agreeNorm}\subref{t:agree3}, 
the correspondence with the known disease status is poorer
than the one obtained with the ranking-based analysis.
In all other cases the MNorm-mix model selected the single component 
homogeneous model as best fitting. 
However, if one forces the model to be fitted as heterogeneous 
then a variable number of groups is selected, ranging from $4$ to $7$. 
Indeed, the best classification that one can obtain with a MNorm-mix fitted to the
rescaled log-transformed absorbances 
of all the $13$ Hum
has a very good agreement with the three disease subgroups, as shown in Table~\ref{t:agreeNorm}\subref{t:agree4}. 
However, we stress that this model is not selected as the best fitting in terms of BIC and yields a more scattered clustering. Moreover, this model requires 117 parameters and hence it is less parsimonious and would be more difficult to interpret
than the best fitting mixture for ranked data.

\begin{table}[]
\caption{Correspondence between the model-based clustering derived by the MLE of the MNorm-mix and the true disease status of the LFPD experimental units:
HD  $=$ healthy,
EBC $=$ diagnosed with early stage breast cancer and
MBC $=$ diagnosed with metastatic breast cancer.}
\label{t:agreeNorm}
\centering
\subfloat[Raw LFPD data with 11 Hum]
{\begin{tabular}{lccc}
& \multicolumn{3}{c}{Group} \\
\cline{2-4}
\multicolumn{1}{c}{Disease Status} & 1 & 2 & 3 \\ 
  \hline
HD  &   7 &   9 &   4 \\ 
EBC &   3 &   2 &  23 \\ 
MBC &   9 &   0 &  10 \\ 
   \hline
\end{tabular}
\label{t:agree3}}\quad
\hspace*{2cm}
\subfloat[Rescaled log-transformed LFPD data with 13 Hum]
{\begin{tabular}{lccccccc}
& \multicolumn{7}{c}{Group} \\
\cline{2-8}
\multicolumn{1}{c}{Disease Status} & 1 & 2 & 3 & 4 & 5 & 6 & 7 \\ 
  \hline
HD &   7 &  11 &   2 &   0 &   0 &   0 &   0 \\ 
  EBC &   0 &   0 &  13 &   1 &  10 &   4 &   0 \\ 
  MBC &   0 &   0 &   3 &   5 &   0 &   9 &   2 \\ 
   \hline
\end{tabular}
\label{t:agree4}}
\end{table}

\section{Concluding remarks and future developments}
\label{concl} 
 
In the present work we presented a novel extension of the popular and widely-used Plackett-Luce model relaxing the standard assumption of forward ranking elicitation and detailed its estimation in the MLE framework. We verified the usefulness of the EPL with a successful application to the real LFPD data set from a bioassay experiment, comparing its performance 
w.r.t. alternative 
and 
more standard
probability distributions for rankings. Specifically, taking into account the heterogeneous origin of the sample units, we considered several parametric models in a mixture model setting.
Inferential results of our mixture modeling approach pointed out a good capability of the absorbance rankings to fit heterogeneous and wilding fluctuating binding data and a good accuracy in discriminating the actual disease status. Interestingly, an almost UM component has been estimated from the data. Differently from previous applications in the literature, where the  UM component was introduced to fit outliers/untypical observations, for the LFPD data such a component does not have the marginal role to model noise in the sample but has a precise interpretation to characterize healthy patients. The utility of the ranking-based analysis for epitope mapping experiments is reinforced by the possibility to partially overcome difficulties related to the choice of the preliminary normalization, needed for the raw quantitative absorbance profiles. Additionally, the fitted model turns out to be more parsimonious than alternative quantitative analyses for the present multivariate setting and exhibits an interesting interpretation,
unaffected by ad-hoc monotone pre-processing transformations of the original raw data.
Hence,
our work 
suggests that 
even when quantitative data are available in a bioassay experiment, 
statistical analysis of the underlying ordinal information 
may provide a useful and more robust tool for the description of the outcomes.
Cluster-specific parameter estimates, characterizing groups of patients, are very useful to construct an epitope mapping profile, i.e., to identify 
protein fragments
whose binding can be 
related to 
the disease development and to detect spots relevant for 
possible 
classification/prediction purposes.
Moreover, the significantly improved fit for the present application obtained with the more general EPL class can be explained with the fact that our proposal accounts for the absence of a natural and \textit{a priori} known reference order of the binding mechanism
and consequently allows to capture the discriminant contribution of all positions. This     
suggests that the understanding of the binding outcomes should not be limited
to the use of the standard forward PL. 

A first natural way to develop further our work consists in implementing the EPL mixture model in a Bayesian framework,
in order to allow the incorporation of pre-experimental information in the analysis. This extension could benefit from the conjugacy of the PL with  
the Gamma prior distribution, already exploited for the Bayesian inference in~\citep{Guiver:Snelson} and~\citep{Caron:Doucet} but restricted to the homogenous population case. Another interesting direction could be that of setting up a flexible framework to integrate
the use of
mixed-type (ordinal and quantitative) data with the possible inclusion of individual covariates.

\section*{Acknowledgements}
We are deeply grateful to Augusto Amici and his research group for providing LFPD data and many insightful discussions on the biological foundations of the experimental outcomes. We also thank
prof. Thomas Brendan Murphy and Isobel Claire Gormley for providing their C code and helpful hints. 

\section*{Appendix}

\begin{center}
\begin{figure}[t]
\centering
\subfloat[][]{
\includegraphics[width=0.32\textwidth]{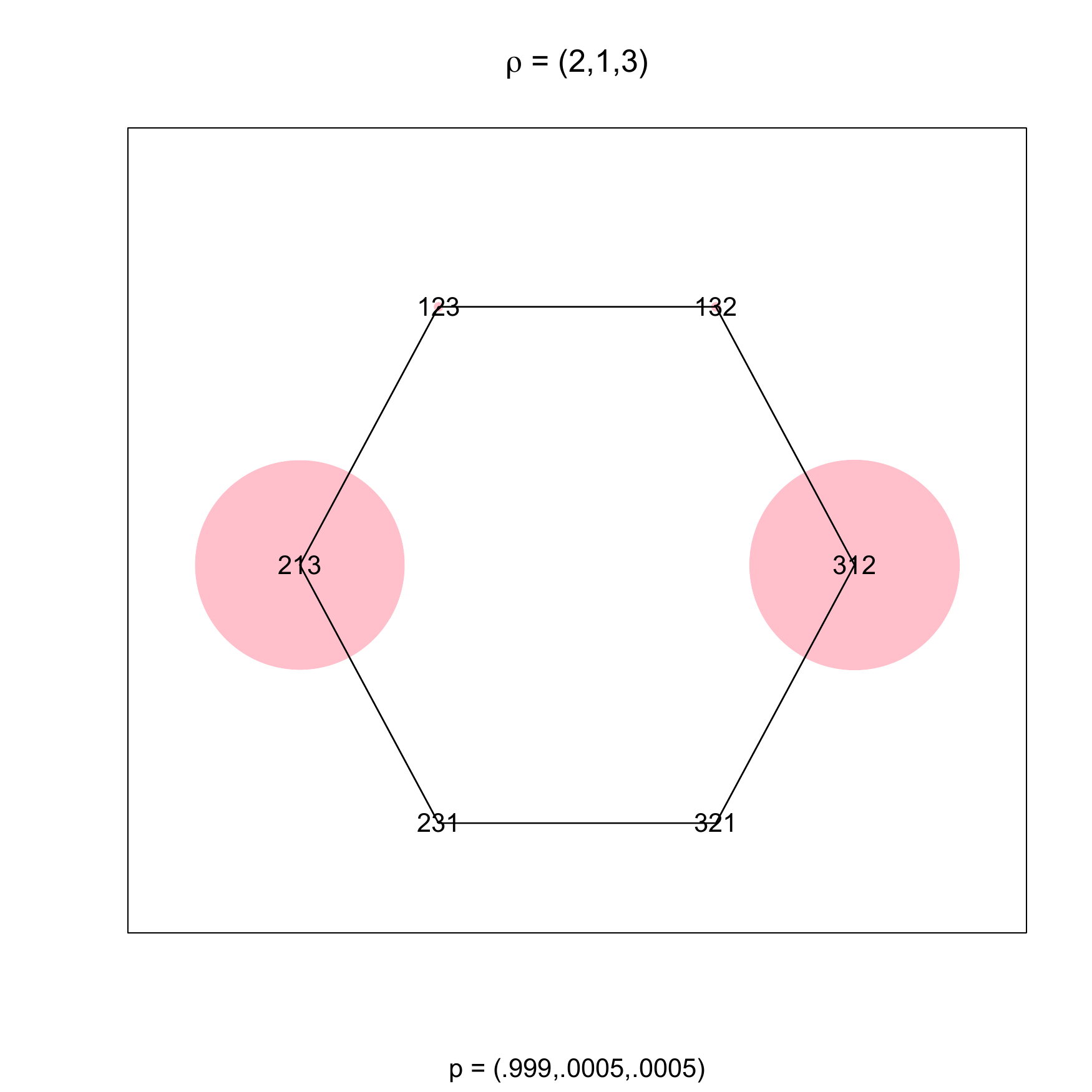} 
\label{fig:IntroEPL1}}
\subfloat[][]{
\includegraphics[width=0.32\textwidth]{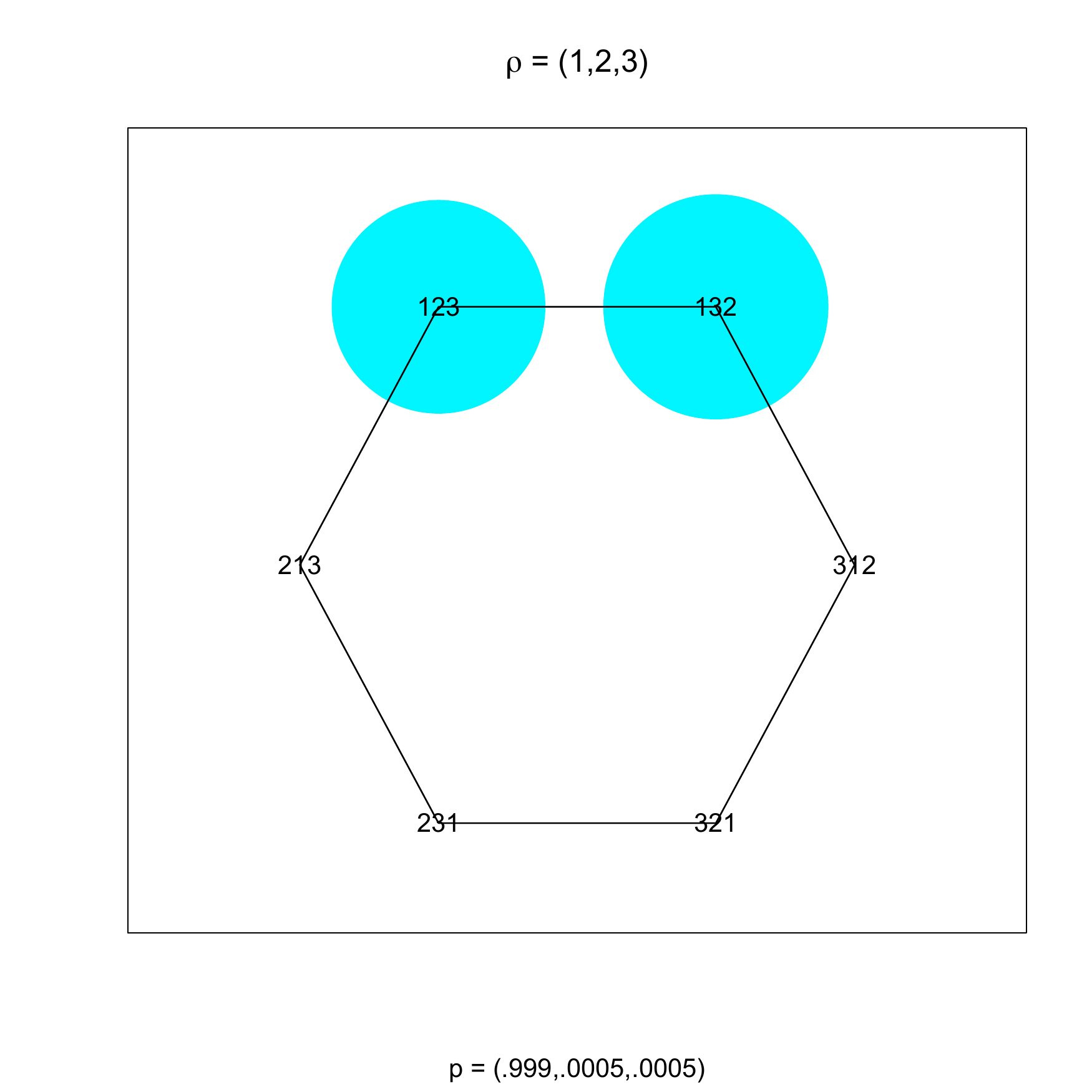}  
\label{fig:IntroEPL2}}
\subfloat[][]{
\includegraphics[width=0.32\textwidth]{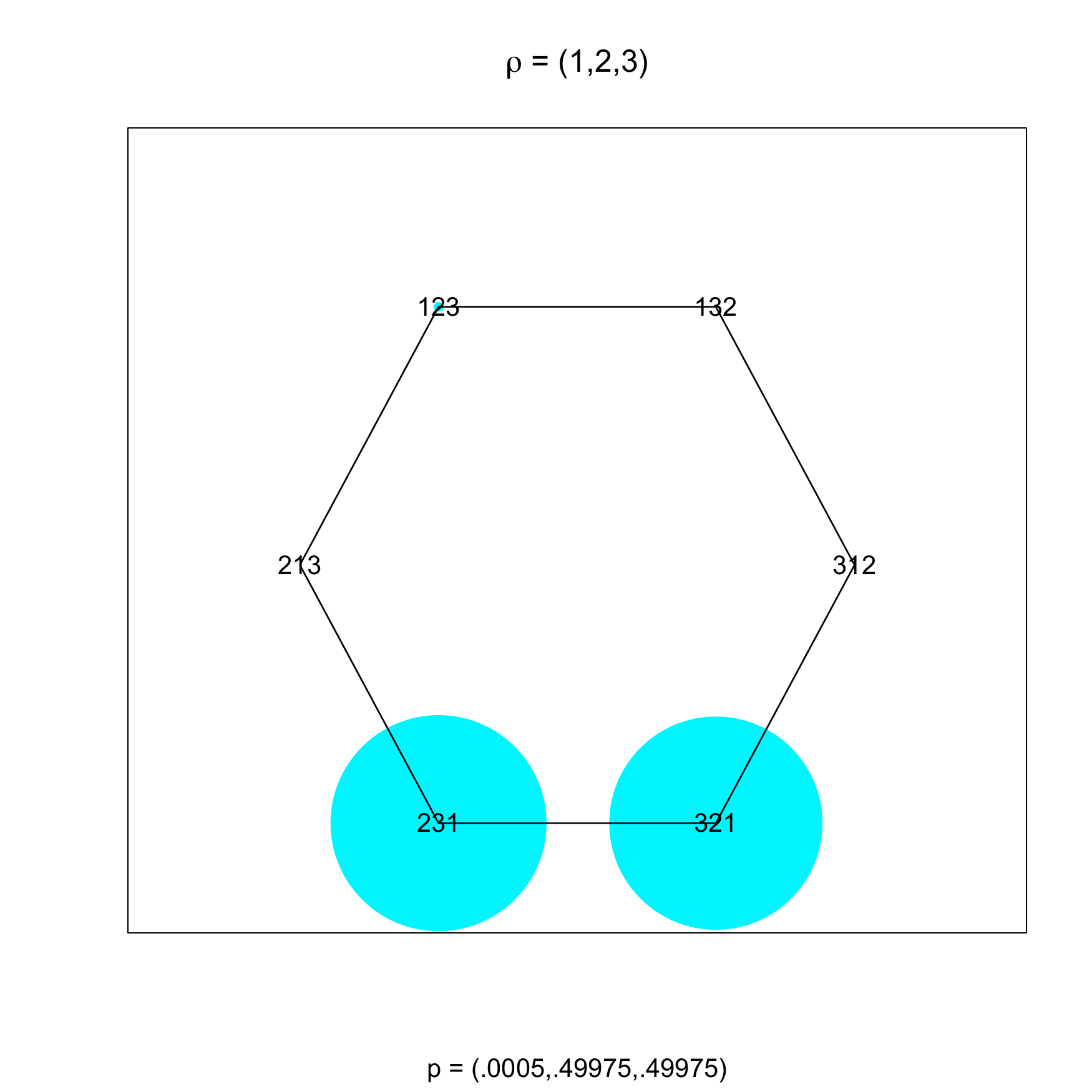}  
\label{fig:IntroEPL3}}
\caption{Examples of EPL (left) and PL (center and right) distribution functions on random orderings.}
\label{fig:IntroEPL}
\end{figure}
\end{center}

We prove here the presence of distributions on orderings in our novel EPL family which are not members of the canonical PL.
For this purpose, let us remind that the PL implies the \textit{independence of irrelevant alternatives} (IIA), stating that the probability ratio to select an item over another is unaffected by the preferences towards the other alternatives in the choice set, see~\citep{Luce}.
Equivalently, one can say that in a PL the choice probability ratio between two items is constant over all stages as long as such alternatives are both still available. In the $K=3$ case the IIA lemma translates into the following set of conditions on the probabilities $q_{\pi^{-1}}=\PP(\pi^{-1})$ of each possible ordering
\begin{eqnarray}
\frac{q_{(1,2,3)}}{q_{(1,3,2)}}&=&\frac{q_{(2,1,3)}+q_{(2,3,1)}}{q_{(3,1,2)}+q_{(3,2,1)}} \notag \\
\frac{q_{(2,1,3)}}{q_{(2,3,1)}}&=&\frac{q_{(1,2,3)}+q_{(1,3,2)}}{q_{(3,1,2)}+q_{(3,2,1)}} \label{e:yes} \\
\frac{q_{(3,1,2)}}{q_{(3,2,1)}}&=&\frac{q_{(1,2,3)}+q_{(1,3,2)}}{q_{(2,1,3)}+q_{(2,3,1)}} \notag
\end{eqnarray}
and they have to be simultaneously satisfied for a generic ranking distribution to belong to the forward PL. Now, let 
us
consider the EPL with fixed $\rho=(2,1,3)$ and the generic induced probability function on random orderings given by
\begin{equation}
\label{e:EPLdistr}
\left(
\begin{matrix}
 q_{(1,2,3)}           & q_{(1,3,2)}          & q_{(2,1,3)}          & q_{(2,3,1)}          & q_{(3,1,2)}          &  q_{(3,2,1)} \\[6pt]
 \frac{p_2p_1}{1-p_2} & \frac{p_3p_1}{1-p_3} & \frac{p_1p_2}{1-p_1} & \frac{p_3p_2}{1-p_3} & \frac{p_1p_3}{1-p_1} & \frac{p_2p_3}{1-p_2} 
\end{matrix}
\right).
\end{equation}
Substituting~\eqref{e:EPLdistr} in~\eqref{e:yes} and solving w.r.t. $\underline{p}$ one obtains as unique solution $\underline{p}=(1/3,1/3,1/3)$, meaning that the two model classes can share only the UM. This formally shows what has been hinted at in~\citep{Fligner:Verducci-American} on the possibility to define new ranking models relaxing the forward hypothesis. To give an intuition about the types of ranking distributions that are not covered by the traditional PL, let us consider the EPL with parameter configuration $\rho=(2,1,3)$ and $\underline{p}=(1-2\epsilon,\epsilon,\epsilon)$ where $\epsilon\to0$. The corresponding probability function over the six possible orderings has two 
equally supported  modes on the sequences with item~1 ranked second capturing almost the total mass, as shown in Figure~\ref{fig:IntroEPL}\subref{fig:IntroEPL1}. This represents a distribution that can not be obtained with any parameter specification from the forward PL. In fact, the suitable calibration of the support parameters can lead only to degenerate marginal choices of item 1 for the first and the last rank, see 
Figures~\ref{fig:IntroEPL}\subref{fig:IntroEPL2} and~\ref{fig:IntroEPL}\subref{fig:IntroEPL3}. Therefore, the introduction of the parameter $\rho$ running in the permutation space allows to overcome this asymmetry among ranks.
  
\bibliographystyle{plainnat}
\bibliography{MT-arxiv}

\end{document}